\documentclass{aa}  
\usepackage{graphicx}
\usepackage{txfonts}
\begin{document}
   \title{Antifreeze in the hot core of Orion\thanks{Based on observations carried out with ALMA and the IRAM Plateau de Bure Interferometer. IRAM is supported by INSU/CNRS (France), MPG (Germany) and IGN (Spain).}}
   \subtitle{First detection of ethylene glycol in \object{Orion-KL}}

  \author{N. Brouillet \inst{1,2}
          \and
          D. Despois \inst{1,2}
                    \and
          X.-H.  Lu\inst{3,4}
                    \and
          A. Baudry \inst{1,2}
                    \and
         J. Cernicharo \inst{5}
                    \and
         D. Bockel\'ee-Morvan \inst{6}
                    \and
         J. Crovisier \inst{6}
                    \and
          N. Biver \inst{6}
          }
          
 \offprints{N. Brouillet}

 \institute{ Univ. Bordeaux, LAB, UMR 5804, F-33270 Floirac, France\\
              \email{[brouillet;despois;baudry]@obs.u-bordeaux1.fr}
         \and
            CNRS, LAB, UMR 5804, F-33270 Floirac, France
          \and
          Universit\'e de Toulouse, UPS-OMP, IRAP, Toulouse, France\\
          \email{xingheng.lu@irap.omp.eu}
          \and
        CNRS, IRAP, 9 Av. Colonel Roche, BP 44346, 31028 Toulouse Cedex 4, France
          \and
	   Centro de Astrobiolog\'ia (CSIC-INTA), Ctra de Torrej\'on a Ajalvir, km 4, 28850 Torrej\'on de Ardoz, Madrid, Spain\\
	    \email{[jose.cernicharo]@cab.inta-csic.es}          
          \and  
          LESIA, Observatoire de Paris, CNRS, UPMC, Universit\'e Paris-Diderot, 5 place Jules Janssen, 92195 Meudon, France\\
           \email{[dominique.bockelee;jacques.crovisier;nicolas.biver]@obspm.fr}
             }

   \date{Received 11 July 2014; accepted 5 February 2015}


  \abstract
   {Ices are present in comets and in the mantles of interstellar grains. Their  chemical composition has been indirectly derived by observing molecules released in the gas phase, when comets approach the sun and when ice mantles are sublimated or destroyed, e.g. in the hot cores present in high-mass, star-forming regions. Comparison of these chemical compositions sheds light on the formation of comets and on the evolution of interstellar matter from the molecular cloud to a protoplanetary disk, and it shows, to first order, a good agreement  between the cometary and
interstellar abundances. However, a complex O-bearing organic molecule, ethylene glycol  (CH$_2$OH)$_2$,  seems to depart from this correlation because it was not easily detected in the interstellar medium (Sgr B2) although it proved to be rather abundant with respect to other O-bearing species in comet C/1995 O1 (Hale-Bopp). Ethylene glycol thus appears, together with the closely related molecules glycolaldehyde CH$_2$OHCHO and ethanol  CH$_3$CH$_2$OH, as a key species in the comparison of interstellar and cometary ices as well as in any discussion on the formation of cometary matter. 
}
   { It is important to measure the molecular abundances in various hot cores to see if the observed differences between the interstellar medium and the comets are general. We focus here on the analysis of ethylene glycol in the nearest and best studied hot core-like region, \object{Orion-KL}.}
   {We use ALMA interferometric data because high spatial resolution observations allow us to reduce the line confusion problem with respect to single-dish observations since different molecules are expected to exhibit different spatial distributions. Furthermore, a large spectral bandwidth is needed because many individual transitions are required to securely detect large organic molecules. Confusion and continuum subtraction are major issues and have been handled with care.}
   {We have detected the \textit{aGg'} conformer of ethylene glycol in \object{Orion-KL}. The emission is compact and peaks towards the Hot Core close to the main continuum peak, about 2$\arcsec$ to the south-west; this distribution is notably different from other O-bearing species. Assuming optically thin lines and local thermodynamic equilibrium, we derive a rotational temperature of 145$\pm$30~K and a column density of 4.6$\pm$0.8~10$^{15}~$cm$^{-2}$. The limit on the column density of the \textit{gGg'} conformer is five times lower.}
   {}

   \keywords{Astrochemistry -- ISM: molecules -- Radio lines: ISM -- ISM: individual objects: Orion-KL  -- Comets: general}

   \maketitle
%
\section{Introduction}

The good global correlation between  the chemical abundances in cometary ices and molecular hot cores
\citep[see Fig.3 of][]{Bockelee-Morvan:2000} is a strong argument in 
favour of similar formation processes in these two rather different types of objects. Ion-molecule and surface reactions at a low ($<$ 200~K) temperature are thus invoked  \citep{Herbst:2009} but one may not yet exclude the possibility of a very direct connection implying comets as being "frozen pieces  of the interstellar medium"  as discussed as early as 1975 by Shimizu \citep[see also Despois 1992, and references therein, and][for a recent review]{Mumma:2011}. 

However, at least two 
important problems arise: the D/H ratio \citep[e.g.][]{Ceccarelli:2014}, and the 
crystalline/amorphous ratio in silicates do not seem to match well \citep[e.g.][]{Crovisier:1997}. Ethylene glycol may be a third example of a discrepancy. 
This opens the possibility for a wide range of models, ranging from pure interstellar matter to pure products of nebular chemistry, to explain the formation of comets. Consequently, different models could explain the protosolar nebula and solar system formation \citep[see e.g.][]{Irvine:2000}. This in turn has consequences on 
the way the collapse proceeds in young stellar objects, on physical and chemical evolution and timescales for the collapse and pre-collapse phase and later mixing in the disk due to turbulence \citep{Bockelee-Morvan:2002}.

Ethylene glycol, HOCH$_2$-CH$_2$OH, is a dialcohol, a molecule chemically related to standard alcohol
 (ethanol, CH$_3$-CH$_2$OH). Ethylene glycol is also known as an antifreeze coolant for car engines. Ethylene 
glycol was first found in the interstellar medium towards the Galactic centre along the line of sight to Sgr~B2 despite line blending problems \citep{Hollis:2002,requena-torres:2008}. In comets, it has been identified for the first time in the Comet C/1995 O1 (Hale-Bopp) spectra by \citet{Crovisier:2004}.

The detailed comparison of the abundances of ethylene glycol in the interstellar medium and in comets
reveals a very different situation:

- In Hale-Bopp, ethylene glycol is rather abundant (0.25~\% by number with respect to H$_2$O, or 1~\% by mass), just behind the  main CHO species, methanol CH$_3$OH (2.5~\% by number), and comparable to formaldehyde (0.1--1.1~\% by number). Ethylene glycol has recently been detected in two other comets: C/2012 F6 (Lemmon) and C/2013 R1 (Lovejoy), with similar abundances relative to water \citep[0.24 and 0.33~\%,][]{Biver:2014}. The simpler ethanol is not detected ($<$ 0.1~\% -- 3$\sigma$ limit), nor the companion molecule to ethylene glycol, glycolaldehyde (HOCH$_2$-CHO) ($<$ 0.04~\%) \citep{Crovisier:2004a}.
 
- In the interstellar medium, ethylene glycol is not easy to identify, and until now it has been found towards only a few  sources: Sgr~B2 \citep{Hollis:2002,requena-torres:2008}, NGC~7129 FIRS~2 \citep{Fuente:2014}, and NGC~1333-IRAS2A \citep{Maury:2014}. Along the 
Sgr~B2 line of sight, glycolaldehyde is seen with a similar abundance \citep[within a factor of 2,][]{Hollis:2002}. Ethanol is more abundant by an order of magnitude in Sgr~B2 \citep[][]{Nummelin:2000}, and seen in many sources. 

How can these differences be explained? Is the line of sight towards the 
galactic centre atypical, or are we seeing general differences between 
interstellar and cometary ices? Observations of hot cores, which appear to be the interstellar regions 
most similar to comets, are required to answer these questions.

\object{Orion-KL}  is one of the main targets to search for new molecules in the interstellar medium, because of its molecular wealth and its proximity.  Interferometric maps are absolutely needed because they 
notably reduce the confusion problem due to line blends in the single dish spectra. As shown by 
\citet{Wright:1996} and confirmed by our data \citep{Guelin:2008,Favre:2011a,Peng:2012,Peng:2013,Brouillet:2013}, different sets of molecules peak towards different positions in \object{Orion-KL}, and possibly even do not 
overlap on high (1--2$\arcsec$) spatial resolution maps. A second clear 
advantage of interferometer maps, is the information they provide: if 
the spatial distribution is identical for various lines, they most 
likely arise from the same species and not from a single interloper exhibiting the same frequencies as those of the molecular species under study. This is especially true for ethylene glycol, a molecule with a complex microwave spectrum. 

In this paper, we report the results of our search for the two most stable conformers of ethylene glycol towards \object{Orion-KL}. The spatial distribution of the ethylene glycol emission and the derived physical parameters are presented in Sect. \ref{sec:results}. In Sect. \ref{sec:discussion} we discuss the chemistry in comparison to other interstellar sources and comets. We present the continuum emission map and its variation at nearby frequencies in Sect. \ref{sec:continuum}. Finally our conclusions are summarized in Sect. \ref{sec:conclusion}.

\section{Observations}
We primarily use here the ALMA Science Verification (ALMA-SV) data\footnote{http://almascience.eso.org/almadata/sciver/OrionKLBand6/} taken on 20 January 2012 towards the IRc2 region in Orion. The observations were carried out with 16 x 12~m antennas in the Band 6 frequency range from 213.715 to 246.627~GHz. 20 slightly overlapping spectral windows were observed, each with a 1.875~GHz bandwidth and 3\,840 channels resulting in a 0.488~MHz channel spacing corresponding to a velocity separation of 0.64~km~s$^{-1}$. The observations were centred on coordinates ($\alpha_{J2000}$ = 05$^{h}$35$^{m}$14$\fs$35, $\delta_{J2000}$ = $-$05$\degr$22$\arcmin$35$\farcs$00) with a primary beam of $\sim$27$\arcsec$.

We used the CASA software\footnote{http://casa.nrao.edu} for initial processing, and then we exported visibilities to the GILDAS package\footnote{http://www.iram.fr/IRAMFR/GILDAS} for further analysis. The line maps were cleaned using the CLARK algorithm \citep{Clark:1980}. The synthesized beam is 1.90$\arcsec$ $\times$ 1.40$\arcsec$ with a PA of 170$\degr$ and the brightness temperature to flux density conversion factor is 9~K for 1~Jy per beam. The continuum emission was subtracted in the maps by carefully selecting line-free channels (see Sect. \ref{sec:continuum}).

We also analyzed several observational data sets obtained with the IRAM Plateau de Bure Interferometer \citep[see Table 1 in][]{Favre:2011a}. These observations cover about 2~GHz from 80.5~GHz to~226 GHz with a velocity separation ranging from 0.42~km~s$^{-1}$ to 2.33~km~s$^{-1}$ and a spatial resolution ranging from 1.79$\arcsec$ $\times$ 0.79$\arcsec$ to 7.63$\arcsec$ $\times$ 5.35$\arcsec$, depending on the frequency and the IRAM interferometer configuration. Four data sets correspond to frequencies included in the ALMA-SV data. In particular, the IRAM 223.4--223.9~GHz observations have spectral and spatial resolution and sensitivity similar to those in the ALMA data set for the same spectral range.

\section{Results}
\label{sec:results} 

\subsection{Conformers of ethylene glycol and selection of lines}
\label{sec:lines} 

Ethylene glycol is an asymmetric top molecule with coupled rotation around its C--C bond and its two C--O bonds leading to different spatial arrangements or conformers \citep[see][]{Christen:1995,Christen:2001}. There are six  conformers relative to the C--C bond with gauche (G) arrangement of the two ethylene groups and four less stable conformers with anti-arrangement of the CH$_2$ groups. In the G arrangement, the two OH groups adopt the gauche orientation with respect to each other. The notations a and g are used for the anti (a) and gauche (g) orientations of the OH groups with respect to the two C--O bonds. When the OH group provides the hydrogen intramolecular bonding, the gauche orientation is noted g'. There are only two conformers with intramolecular bonding, \textit{aGg'} and \textit{gGg'}, and \citet{Muller:2004} showed that the latter lies about 200 cm$^{-1}$ or about 290~K higher in energy than \textit{aGg'}. For these two conformers, tunneling is observed between two equivalent equilibrium configurations and splits each rotational level into two distinct states designated v = 0 and v = 1 \citep[e.g.][]{Christen:2001}. The v = 1 state is around 7 and 1.4~GHz higher than the v = 0 state for the \textit{aGg'} and \textit{gGg'} conformers, respectively \citep{Christen:2003}.

The rotation-tunnelling microwave spectra of the \textit{aGg'} and \textit{gGg'} conformers were accurately determined in the 54--370~GHz \citep{Christen:2003} and 77--579~GHz \citep{Muller:2004} ranges, respectively. The results from these works and earlier works are now available from the Cologne database for molecular spectroscopy \citep[CDMS\footnote{http://www.astro.uni-koeln.de/cdms},][] {Muller:2001,Muller:2005} and can be used for reliable identification of the ethylene glycol transitions. Earlier ethylene glycol line predictions were used to assign to the \textit{aGg'} conformer five features observed by \citet{Hollis:2002} in Sgr B2; two other features were also assigned to the same conformer in the same source by \citet{requena-torres:2008}. The CDMS was used to identify more than 10~\textit{aGg'} conformer lines in the Hale-Bopp comet \citep{Crovisier:2004}. We note that several ethylene glycol transitions observed in the direction of Sgr~B2 are blended, while transitions are well 'isolated' towards Hale-Bopp.  

We used the most recent CDMS data to search for transitions for the \textit{aGg'} and \textit{gGg'} conformers in the ALMA-SV data cube. From line intensity predictions at 120\,K, a typical temperature for the region \citep[e.g.][]{Favre:2011a}, we have examined the strongest lines among the ethylene glycol transitions expected in the different spectral bands. We produced synthetic spectra for both conformers (see Sect. \ref{sec:parameters} and \ref{sec:gGg}). We also derived synthetic spectra for higher temperatures of the \textit{gGg'} conformer to investigate the presence of the \textit{gGg'} conformer in hotter gas because \textit{gGg'} is higher in energy than \textit{aGg'} (see above).

\begin{table*}
\begin{minipage}[t]{17cm}
\caption{Ethylene glycol transitions (\textit{aGg'} conformer) not blended or only partially blended in the ALMA dataset.}             
\label{table.freq-eg}      
\centering                          
\small\addtolength{\tabcolsep}{-1pt} 
\renewcommand{\footnoterule}{}  
\begin{tabular}{l l c c c c  }        
\hline\hline                 
Frequency \footnote{Transitions used for the rotational diagram (see Fig. \ref{dr-eth-eg}) are indicated by a star.}	&	Quantum numbers  \footnote{The quantum numbers are \textit{J}(\textit{K$_\mathrm{a}$},\textit{K$_\mathrm{c}$}) as defined in the CDMS database where v is a state number (see Sect. \ref{sec:lines}).}	&	\it{S}$\mu$$^2$ \footnote{$\it{S}$ is the line strength and $\mu$ the dipole moment.} &	$\log $(\it{A}$_\mathrm{ul}$) \footnote{See Sect. \ref{sec:parameters} for the definition of the parameters.} &	\it{E}$_\mathrm{u}$ $^d$ &	\it{g}$_\mathrm{u}$ $^d$	\\
(MHz)  &  &  (D$^2$) &   &(K) &    \\
\hline                        
 
214\,526.41410	&	19(11, 8) v = 1 -- 19(10, 9) v  = 1	&	50.7	&	 $-$4.67073	&	153.1	&	273	\\
214\,526.41470	&	19(11, 9) v = 1 -- 19(10,10) v  = 1	&	65.2	&	 $-$4.67067	&	153.1	&	351	\\
214\,712.96650	&	21(5,16) v = 0 -- 20(5,15) v  = 1	&	732.5	&	$-$3.66141	&	127.1	&	387	\\
214\,808.01940	&	23(2,22) v = 0 -- 22(2,21) v  = 1	&	685.7	&	$-$3.61897	&	132.4	&	329	\\
214\,829.92520	&	23(1,22) v = 0 -- 22(1,21) v= 1	&	875.5	&	$-$3.62187	&	132.4	&	423	\\
216\,685.81470$^*$	&	21(3,19) v = 1 -- 20(3,18) v = 0	&	663.4	&	$-$3.69253	&	117.3	&	387	\\
216\,826.11240	&	20(5,15) v = 1 -- 19(5,14) v = 0	&	557.3	&	$-$3.74666	&	116.8	&	369	\\
217\,139.72350	&	21(4,17) v = 0 -- 20(4,16) v = 1	&	786.8	&	$-$3.61568	&	123.9	&	387	\\
217\,449.99470	&	24(1,24) v = 1 -- 23(1,23) v = 1	&	928.6	&	$-$3.59857	&	136.4	&	441	\\
217\,450.27020$^*$	&	24(0,24) v = 0 -- 23(0,23) v = 1	&	722.2	&	$-$3.59863	&	136.4	&	343	\\
217\,587.54780$^*$	&	21(2,19) v = 1 -- 20(2,18) v = 0	&	666.2	&	$-$3.57614	&	117.2	&	301	\\
219\,089.72000	&	22(10,13) v = 0 -- 21(10,12) ) v = 1	&	670.7	&	$-$3.69311	&	173.5	&	405	\\
219\,089.72790	&	22(10,12) v = 0 -- 21(10,11) v = 1	&	521.6	&	$-$3.69317	&	173.5	&	315	\\
219\,384.91030$^*$	 &	26(11,16) v = 1 -- 26(10,16) v = 0	& 15.8	& $-$5.28218	& 232.4	& 371 \\
219\,385.17780$^*$	 &	22( 9,14) v = 0 -- 21( 9,13) v = 1	& 703.8	& $-$3.67045	& 164.3	& 405 \\
219\,385.32400$^*$	 &	26(11,15) v = 1 -- 26(10,17) v = 0	& 20.3	& $-$5.28213	& 232.4	& 477 \\
219\,385.42560$^*$	 &	22( 9,13) v = 0 -- 21( 9,12) v = 1	& 547.3	& $-$3.67051	& 164.3	& 315 \\
221\,007.82310	&	21(4,18) v = 1 -- 20(4,17) v = 0	&	818.3	&	$-$3.57562	&	122.1	&	387	\\
221\,038.79970$^*$	&	22(6,17) v = 0 -- 21(6,16) v = 1	&	771.6	&	$-$3.62074	&	142.6	&	405	\\
221\,100.31520	&	22(5,18) v = 0 -- 21(5,17) v = 1	&	762.5	&	$-$3.62553	&	137.4	&	405	\\
222\,054.23240$^*$	 &	21(10,12) v = 1 -- 20(10,11) v = 0	& 624.5	& $-$3.68686	& 163.0	& 387 \\
222\,054.23580$^*$	 &	21(10,11) v = 1 -- 20(10,10) v = 0	& 485.8	& $-$3.68681	& 163.0	& 301 \\
222\,348.60040$^*$	&	23(1,23) v = 1 -- 22(1,22) v = 0	&	889.5	&	$-$3.57018	&	126.0	&	423	\\
222\,349.15090$^*$	&	23(0,23) v = 1 -- 22(0,22) v = 0	&	692.0	&	$-$3.57003	&	126.0	&	329	\\
223\,741.66380$^*$	&	21(6,16) v = 1 -- 20(6,15) v = 0	&	738.2	&	$-$3.60437	&	132.0	&	387	\\
226\,643.30070$^*$	&	25(1,25) v = 0 -- 24(1,24) v = 1	&	752.5	&	$-$3.54420	&	147.7	&	357	\\
226\,643.45630$^*$	&	25(0,25) v = 0 -- 24(0,24) v = 1	&	967.6	&	$-$3.54415	&	147.7	&	459	\\
228\,602.83320	&	23(16,7) v = 0 -- 22(16,6) v = 1	&	456.4	&	$-$3.82381	&	261.3	&	423	\\
228\,602.83320	&	23(16,8) v = 0 -- 22(16,7) v = 1	&	354.9	&	$-$3.82387	&	261.3	&	329	\\
228\,752.77650$^*$\footnote{Not blended transition.}	&	23(2,22) v = 1 -- 22(2,21) v = 0	&	875.2	&	$-$3.54020	&	132.7	&	423	\\
229\,233.71610	&	23(11,13) v = 0 -- 22(11,12) v = 1	&	530.2	&	$-$3.64599	&	195.1	&	329	\\
229\,233.71660	&	23(11,12) v = 0 -- 22(11,11) v = 1	&	681.8	&	$-$3.64594	&	195.1	&	423	\\
231\,127.40080$^*$	&	23(7,16) v = 0 -- 22(7,15) v = 1	&	801.1	&	$-$3.56515	&	160.2	&	423	\\
231\,524.03310$^*$	 &	23( 6,18) v = 0 -- 22( 6,17) v= 1	& 618.1	& $-$3.56638	& 154.1	& 329 \\
231\,564.00490	&	24(1,24) v = 1 -- 23(1,23) v = 0	&	722.1	&	$-$3.51672	&	136.8	&	343	\\
231\,564.31950	&	24(0,24) v = 1 -- 23(0,23) v = 0	&	928.6	&	$-$3.51666	&	136.8	&	441	\\
232\,095.73820	&	22(12,11) v = 1 -- 21(12,10) v = 0	&	462.3	&	$-$3.67043	&	195.4	&	315	\\
232\,095.73830	&	22(12,10) v = 1 -- 21(12,9) v = 0	&	594.5	&	$-$3.67038	&	195.4	&	405	\\
232\,349.45090	&	22(20,2) v = 1 -- 21(20,1) v = 0	&	146.9	&	$-$4.27619	&	320.9	&	405	\\
232\,349.45090	&	22(20,3) v = 1 -- 21(20,2) v = 0	&	114.2	&	$-$4.27614	&	320.9	&	315	\\
232\,350.05940	&	22(10,13) v = 1 -- 21(10,12) v = 0	&	522.1	&	$-$3.61615	&	173.8	&	315	\\
232\,350.06810	&	22(10,12) v = 1 -- 21(10,11) v = 0	&	671.4	&	$-$3.61610	&	173.8	&	405	\\
232\,881.53340	&	23(6,17) v = 0 -- 22(6,16) v = 1	&	750.3	&	$-$3.58376	&	154.2	&	423	\\
235\,304.05030$^*$	&	22(6,16) v = 1 -- 21(6,15) v = 0	&	773.7	&	$-$3.53805	&	143.1	&	405	\\
235\,441.43720	&	10(7,4) v = 0 -- 9(6,3) v = 0	&	56.6	&	$-$4.34165	&	51.1	&	189	\\
235\,441.48820	&	10(7,3) v = 0 -- 9(6,4) v = 0	&	44.1	&	$-$4.34161	&	51.1	&	147	\\
235\,442.34320	&	10(7,4) v = 1 -- 9(6,3) v = 1	&	42.1	&	$-$4.36183	&	51.4	&	147	\\
235\,442.39560	&	10(7,3) v = 1 -- 9(6,4) v = 1	&	54.1 &	$-$4.36177	&	51.4	&	189	\\
235\,471.89630	&	8(8,0) v = 0 -- 7(7,1) v = 0	&	49.0	&	$-$4.20372	&	49.1	&	119	\\
235\,471.89630	&	8(8,1) v = 0 -- 7(7,0) v = 0	&	63.0	&	$-$4.20376	&	49.1	&	153	\\
235\,834.23970$^*$	&	26(1,26) v = 0 -- 25(1,25) v = 1	&	1\,006.5	&	$-$3.49194	&	159.3	&	477	\\
235\,834.32720$^*$	&	26(0,26) v = 0 -- 25(0,25) v = 1	&	782.7	&	$-$3.49199	&	159.3	&	371	\\
239\,605.33240$^*$	&	24(11,14) v = 0 -- 23(11,13) v = 1	&	728.5	&	$-$3.57758	&	206.9	&	441	\\
239\,605.33370$^*$	&	24(11,13) v = 0 -- 23(11,12) v = 1	&	566.7	&	$-$3.57753	&	206.9	&	343	\\
241\,545.26260	&	24(7,18) v = 0 -- 23(7,17) v = 1	&	841.5	&	$-$3.50446	&	172.1	&	441	\\
242\,947.99050$^*$	&	23(9,15) v = 1 -- 22(9,14) v = 0	&	749.2	&	$-$3.52928	&	175.9	&	423	\\
242\,948.59120$^*$	&	23(9,14) v = 1 -- 22(9,13) v = 0	&	582.8	&	$-$3.52923	&	175.9	&	329	\\
243\,259.73980	&	24(6,18) v = 0 -- 23(6,17) v = 0	&	324.4	&	$-$3.80007	&	166.2	&	343	\\
244\,685.14770	&	24(2,22) v = 1 -- 23(2,21) v = 1	&	939.2	&	$-$3.43992	&	150.1	&	441	\\
 
\hline                                   
\end{tabular}
\end{minipage}
\end{table*}

%
\subsection{Spatial distribution and line profile of the \textit{aGg'} conformer of ethylene glycol }
\label{sec:spatial} 

\begin{figure}
   \centering
  \includegraphics[width=8cm]{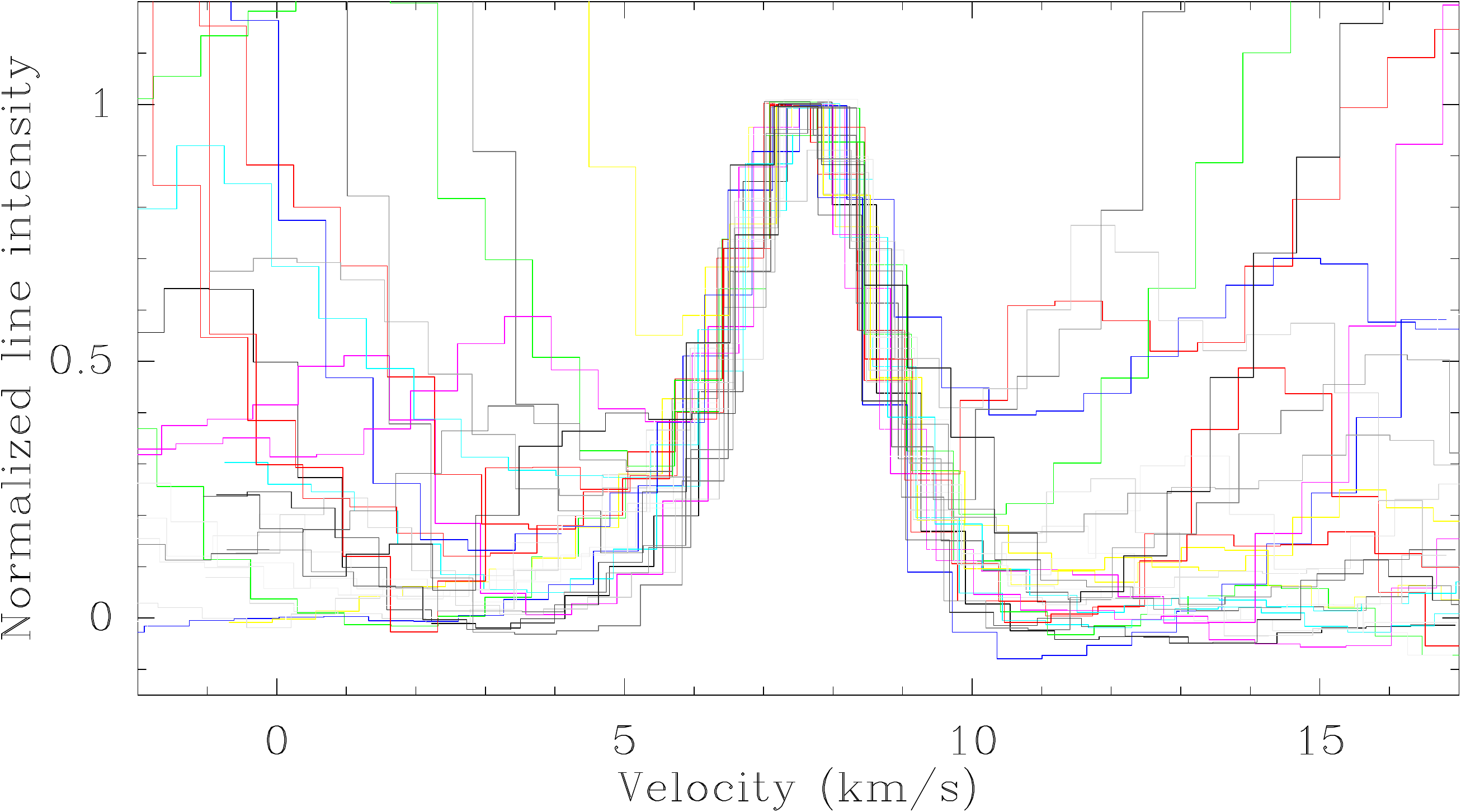}
   \caption{Superposition of the unblended and partially blended ethylene glycol lines listed in Table  \ref{table.freq-eg}. The lines corresponding to close pairs of transitions separated by 0.5--1~MHz are not included in this plot as their profile is broadened. The spectra are towards the ethylene glycol peak ($\alpha_{J2000}$ = 05$^{h}$35$^{m}$14$\fs$47, $\delta_{J2000}$ = $-$05$\degr$22$\arcmin$33$\farcs$17) and are normalized to 1. }
         \label{line-profile}
 \end{figure}
 
 \begin{figure}
   \centering
  \includegraphics[width=8cm]{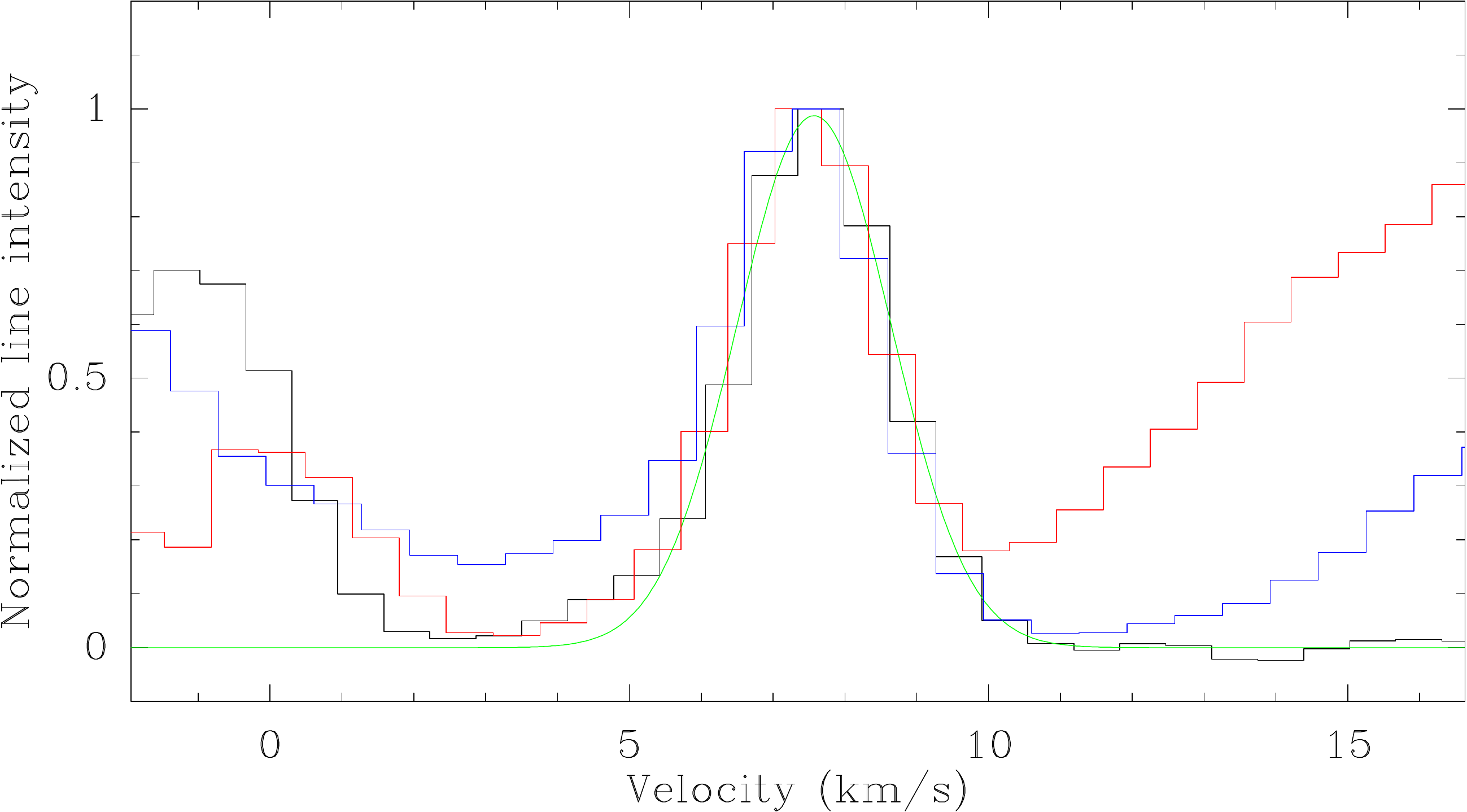}
   \caption{Superposition of the 228\,752~MHz transition spectrum at the ethylene glycol peak (in black) with the average spectrum of 11 transitions that have no blends near 5 km~s$^{-1}$ (in red) and the average spectrum of 14 transitions that have no blends near 10 km~s$^{-1}$ (in blue). All spectra are normalized to 1. A Gaussian fit to the 228\,752~MHz line is shown in green. A faint wing is visible at lower velocities.}
         \label{ailes}
 \end{figure}

\begin{figure}
   \centering
  \includegraphics[width=8cm]{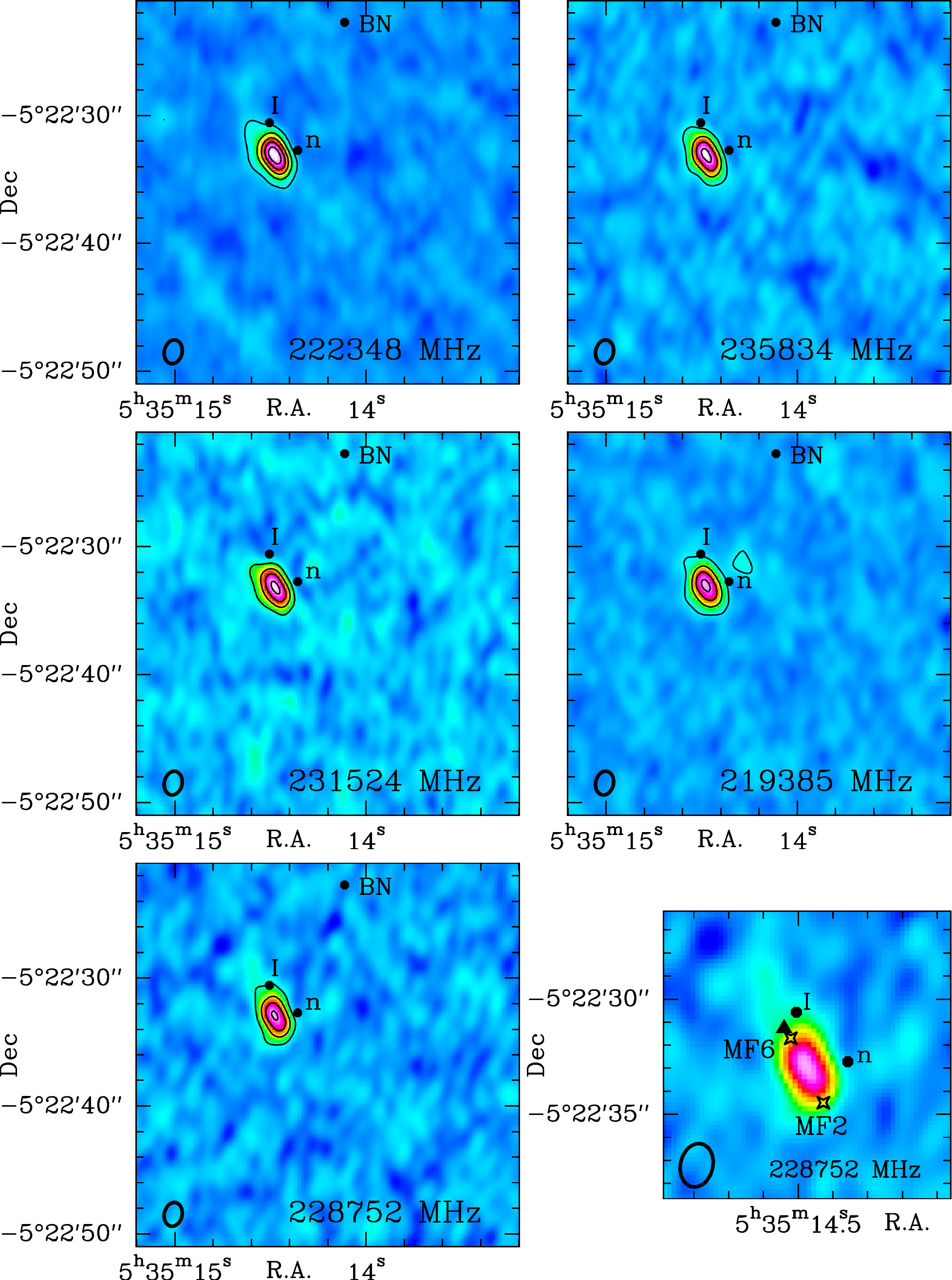}
   \caption{ ALMA ethylene glycol integrated intensity maps of the five less blended lines. The frequency of the transition is indicated at the bottom of the map and the first contour level corresponds to 4$\sigma$. The beam shown in the bottom left corner is 1.9$\arcsec$ $\times$ 1.39$\arcsec$. The BN object position is ($\alpha_{J2000}$ = 05$^{h}$35$^{m}$14$\fs$11, $\delta_{J2000}$ = $-$05$\degr$22$\arcmin$22$\farcs$72), the radio source I position is ($\alpha_{J2000}$ = 05$^{h}$35$^{m}$14$\fs$51, $\delta_{J2000}$ = $-$05$\degr$22$\arcmin$30$\farcs$57), and the IR source n position is ($\alpha_{J2000}$ = 05$^{h}$35$^{m}$14$\fs$36, $\delta_{J2000}$ = $-$05$\degr$22$\arcmin$32$\farcs$72) \citep{Goddi:2011a}. The map to the lower right is a blow up of the 228\,752~MHz map. The triangle marks the position of the continuum emission peak (see map in Fig. \ref{continuum-maps}) and the stars mark the MF2 and MF6 methyl formate positions of \citet{Favre:2011a}. The ethylene glycol emission peaks at ($\alpha_{J2000}$ = 05$^{h}$35$^{m}$14$\fs$47, $\delta_{J2000}$ = $-$05$\degr$22$\arcmin$33$\farcs$17). 
   The lines are integrated over the velocity range 5--10~km~s$^{-1}$ for the 222\,348~MHz transition, 5.5--10~km~s$^{-1}$ for 235\,834~MHz, 7--10~km~s$^{-1}$ for 231\,524~MHz, 6--9.5~km~s$^{-1}$ for 219\,385~MHz and 4--11~km~s$^{-1}$ for 228\,752~MHz. The level step and first contour are 6 and 2.8~K~km~s$^{-1}$ for the 222\,348~MHz transition, 7.6 and 3.6~K~km~s$^{-1}$ for 235\,834~MHz,  2.5 and 1.7~K~km~s$^{-1}$ for 231\,524~MHz, 5 and 1.9~K~km~s$^{-1}$ for 219\,385~MHz and  4.4 and 2.9~K~km~s$^{-1}$ for 228\,752~MHz.}
         \label{cartes-EG}
 \end{figure}

\begin{figure}
   \centering
  \includegraphics[width=8cm]{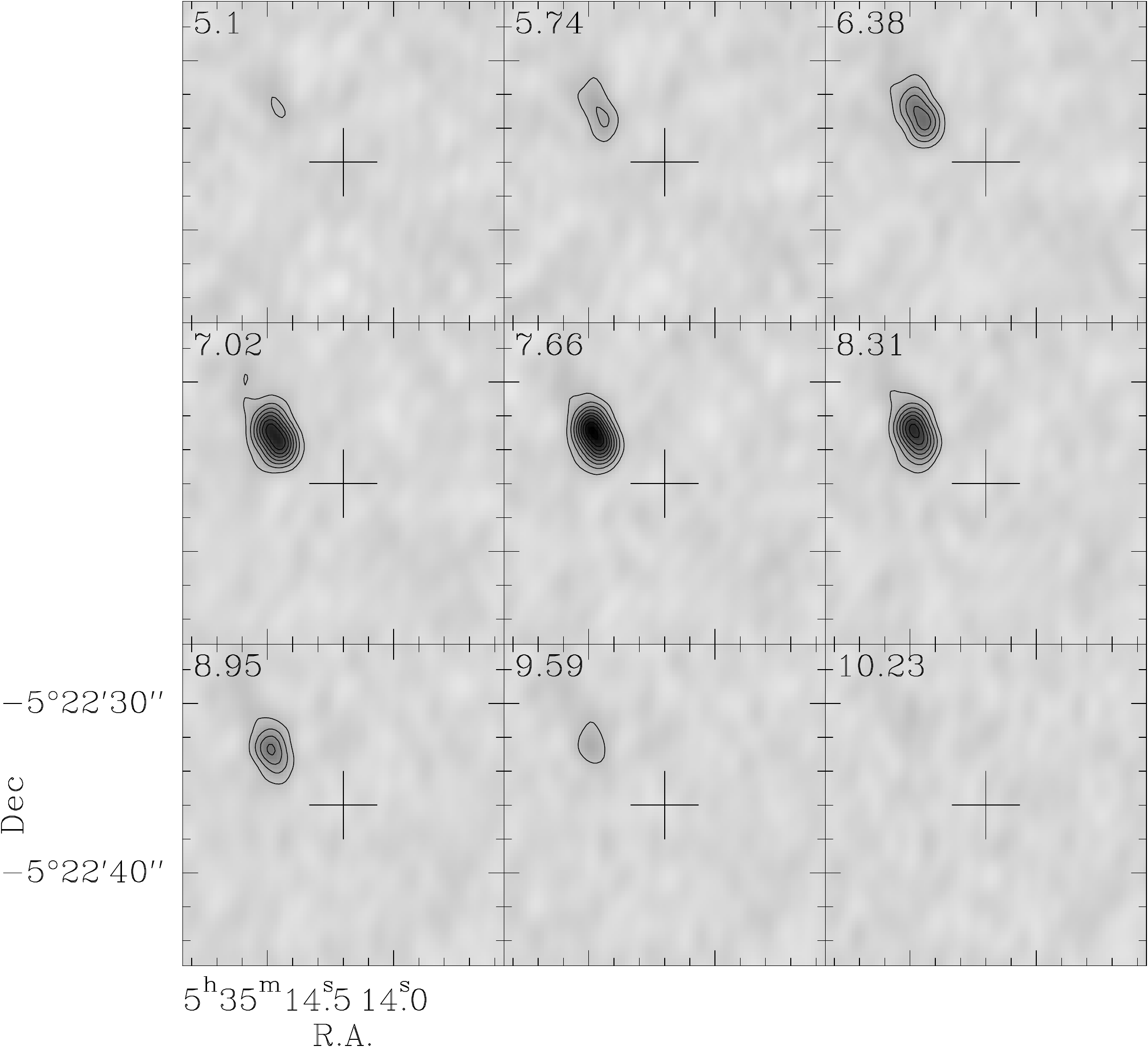}
   \caption{ALMA channel maps of the ethylene glycol 228\,752~MHz transition which has no apparent contamination. The level step and first contour are 0.44~K. The LSR velocity in km~s$^{-1}$ is indicated in the upper left corner of each map.}
         \label{channel-EG-228752}
 \end{figure}

In the ALMA dataset, most ethylene glycol lines suffer mild to heavy blending. Table  \ref{table.freq-eg} lists 57 transitions that are mildly blended and a noncontaminated transition lying at 228\,752~MHz.
All these transitions present a peak in their spatial distributions at ($\alpha_{J2000}$ = 05$^{h}$35$^{m}$14$\fs$47, $\delta_{J2000}$ = $-$05$\degr$22$\arcmin$33$\farcs$17). In Fig. \ref{line-profile} we plotted a superposition of the spectra of the partially blended ethylene glycol lines towards this position, excluding line doublets with a difference in frequency of $\sim$ 0.5--1~MHz. We normalized the spectra to a peak intensity of 1. All line profiles are similar in the 5 to 10~km~s$^{-1}$ velocity range. They can be fitted with a Gaussian centred at a velocity of 7.6~km~s$^{-1}$ and a line width of  2.3~km~s$^{-1}$. We have averaged the transitions that exhibit no blends around 5 km~s$^{-1}$ on one hand, and the transitions that exhibit no blends around 10 km~s$^{-1}$ on the other hand.  Figure \ref{ailes} shows the superposition of these two spectra together with the 228\,752~MHz transition spectrum. The resulting line profile of the unblended parts fits perfectly the profile of the 228\,752 MHz transition. The Gaussian fit shows a faint excess around 5~km~s$^{-1}$. This excess is significant as it is above the noise of the 228\,752 MHz individual spectrum. Furthermore it appears on both profiles, which use different transitions, shown in Fig. \ref{ailes}.

The spatial distribution of ethylene glycol in \object{Orion-KL} is shown in Fig. \ref{cartes-EG}, which presents the intensity maps integrated over the noncontaminated channels of the five best lines (transitions at 219\,385~MHz, 222\,348~MHz, 228\,752~MHz, 231\,524~MHz, and 235\,834~MHz). The 228\,752~MHz transition appears to be the only one free of contamination, i.e. without any blending line above the noise level;  the channel maps are shown in Fig. \ref{channel-EG-228752}. 

The \textit{aGg'} conformer distribution is compact and appears to be slightly bigger than the beam. We find an upper limit for the deconvolved size of 2.4$\arcsec$$\times$ 1.1$\arcsec$. The emission peak is located $\sim$ 2$\arcsec$ to the south-west of the peak of continuum emission (see map in lower right panel of Fig. \ref{cartes-EG} and Fig. \ref{continuum-maps} map); it lies between the methyl formate MF2 and MF6 peaks of \citet{Favre:2011a}. 
The line profile is similar to the methyl formate and ethanol line profiles at the same position, centred at a velocity of $\sim$7.6~km~s$^{-1}$ and with a faint wing around 5~km~s$^{-1}$. A detailed comparison of these O-bearing molecules will be performed in a forthcoming paper.
 

\subsection{Spectral confusion}
\label{sec:confusion} 

The spectral line density in Orion is high \citep[e.g. 8.6 lines per 100~MHz in the IRAM 30m survey of][]{Tercero:2010}, comparable to that in Sgr~B2  \citep[6--10 lines per 100~MHz,][]{Friedel:2004,Halfen:2006} for high-sensitivity observations.

\cite{Snyder:2005} have given the essential criteria for establishing the identification of a new interstellar molecule. Even if ethylene glycol has already been detected in Sgr~B2 and two other interstellar sources, its identification is not obvious because of the weakness of the lines and spectral confusion. The different criteria are:

- The rest frequencies must be known with a high degree of accuracy from laboratory measurements.

- The central velocity and line width must agree for all detected transitions.

- Blended lines must be separated by at least half the line width (FWHM) of the weaker line, to be considered as resolved. 

- Beam dilution must be taken into account when comparing observations made at different frequencies.

- The relative intensities of the different transitions must be consistent (when, as we suppose it is the case here, an excitation close to local thermodynamic equilibrium is likely).

- All transitions with intensity predictions leading to detectable signal levels must be present.

For interferometric observations, another criterion can be added: the coherence of the spatial distribution. Indeed the spatial distribution of the different transitions must be similar for similar excitation conditions (e.g. for a similar upper level energy). The probability of chance alignment in frequency in one spectrum may be high, but a chance alignment for all spectra in an extended map is unlikely. As most of the molecules display different distributions on a high-resolution scale, confusion will not produce similar maps for every transition. Requiring a single interloper for different transitions makes a misidentification much less probable.

Detailed study of the ethylene glycol spectroscopy is relatively recent and we can be confident that the rest frequencies are accurate (see Sect. \ref{sec:lines}). In our study of \object{Orion-KL} , there is a frequency agreement among all the partially blended transitions: the line centre velocities are in good agreement within the uncertainties (7.6$\pm$0.2~km~s$^{-1}$) as well as the line widths and the general profiles (see Fig. \ref{line-profile}). In addition, the noncontaminated parts of these lines display the same spatial distribution (see Fig. \ref{cartes-EG}). The lines that we consider partially blended with others are separated by at least their half-maximum intensity. The relative intensities of the transitions are consistent: no expected strong lines are missing and a rotational temperature can be estimated based on these transitions (see Sect. \ref{sec:parameters}). 

\cite{Halfen:2006} add that the evidence for emission from a given molecule must be over a substantial frequency range and that an extensive, self-consistent data set is necessary to identify large organic species in the interstellar gas. Indeed, our initial search for ethylene glycol in \object{Orion-KL}  based on limited spectral ranges with the IRAM Plateau de Bure Interferometer (PdBI), was not convincing enough. The ALMA Science Verification data are similar to our 223~GHz PdBI data for their sensitivity and their spectral and spatial resolution but they cover 32.9~GHz instead of 0.5~GHz. Using all the PdBI  data sets \citep{Favre:2011a}, which cover $\sim$2~GHz, we have identified seven blended lines coinciding with \textit{aGg'} conformer transitions\footnote{Frequencies of the PdBI ethylene glycol blended lines: 105.7010~GHz, 105.8347~GHz, 203.4309~GHz, 203.4533~GHz, 223.4589~GHz, 223.5205~GHz, 223.7417~GHz.}. However, we have discarded these lines from the present analysis because of line blending problems and because of different spatial and spectral resolution used in our PdBI data.

%
\subsection{Physical parameters of the \textit{aGg'} conformer of ethylene glycol}
\label{sec:parameters} 

Assuming optically thin lines and local thermodynamic equilibrium, we have estimated the rotational temperature  and the column density of ethylene glycol using the equation \citep[e.g.][]{Goldsmith:1999}:  
%
      \begin{equation}
           \rm
     ln(\frac{\it{8\pi k_{B} \nu^{2} W}}{\it{h c^{3}g_\mathrm{u}A_\mathrm{ul} }}) = 
                      ln(\frac{\it{N}}{Q(T)}) - \frac{\it{E_\mathrm{u}}}{\it{k_{B}T}},\\                     
       \end{equation}
where $W$ is the integrated line intensity (K~km~s$^{-1}$), $\nu$ the line frequency (Hz), $A_\mathrm{ul}$ the Einstein coefficient for spontaneous emission (s$^{-1}$), $g_\mathrm{u}$ the degeneracy of the upper state, $N$ the total column density, $Q(T)$ the partition function, $E_\mathrm{u}$ the upper state energy, and $T$ the excitation temperature. 

\begin{figure}
   \centering
  \includegraphics[width=8cm]{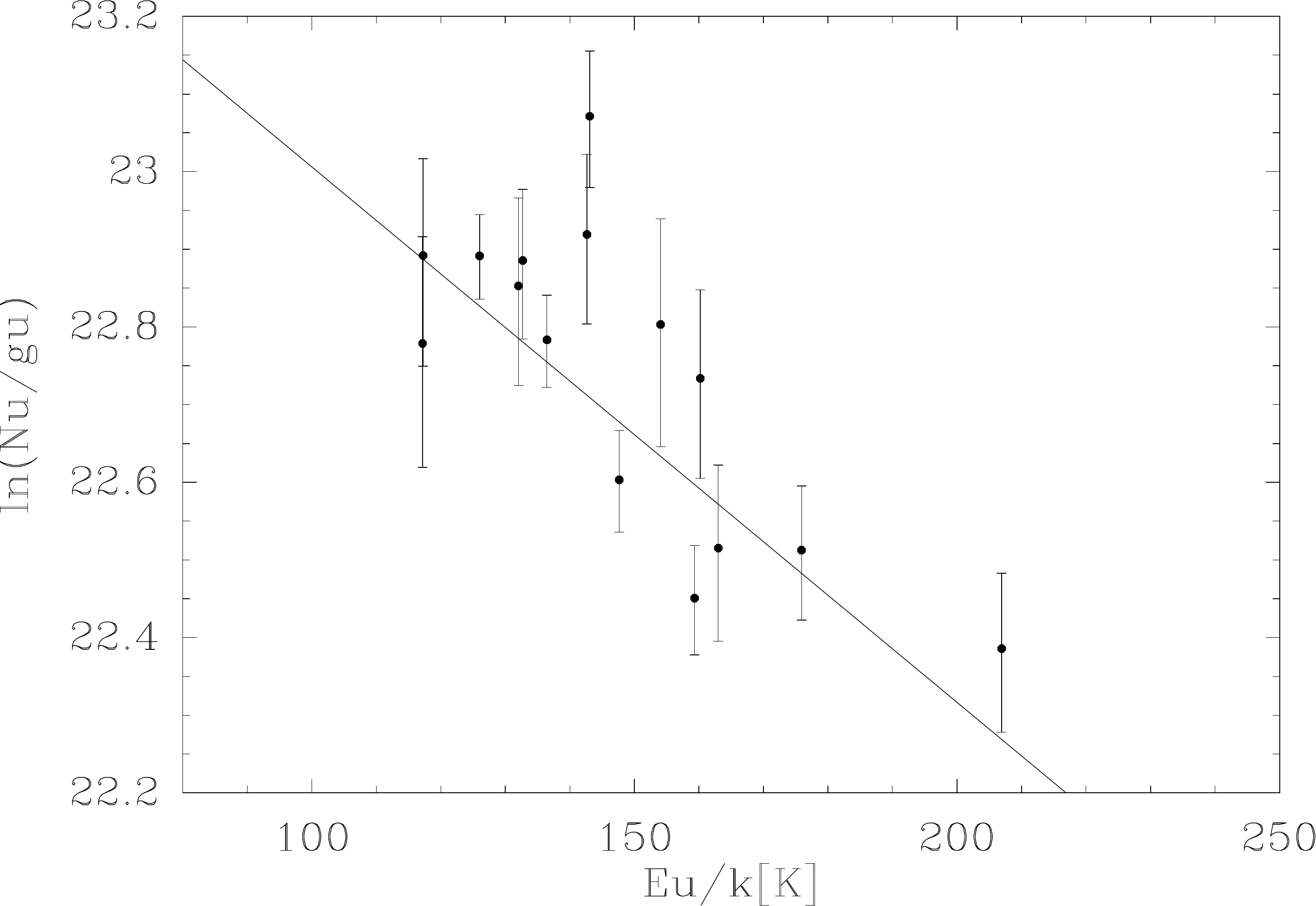}
   \caption{Rotational diagram at the emission peak based on the less-contaminated ethylene glycol lines indicated by a star in Table  \ref{table.freq-eg}. }
         \label{dr-eth-eg}
 \end{figure}

The rotational diagram is obtained at the emission peak from the less-contaminated ethylene glycol lines to derive reliable physical parameters (see Fig. \ref{dr-eth-eg}). Taking the partition function \textit{Q}(\textit{T}) = 234\,500 from the nonlinear plot of the CDMS values of \textit{Q} at 145~K, we obtain a rotational temperature of 145$\pm$30~K and derive a total column density averaged over the synthesized beam \textit{N}$\rm_{\textit{aGg'}}$ of 4.6$\pm$0.8~10$^{15}~$cm$^{-2}$.  From the continuum map (see Sect. \ref{sec:continuum}), we derived the beam averaged projected density N$\rm_{H\rm_{2}}$ as in \citet{Favre:2011a}, and we find an ethylene glycol abundance of  $\sim$10$^{-9}$.

\begin{figure}
   \centering
  \includegraphics[width=8cm]{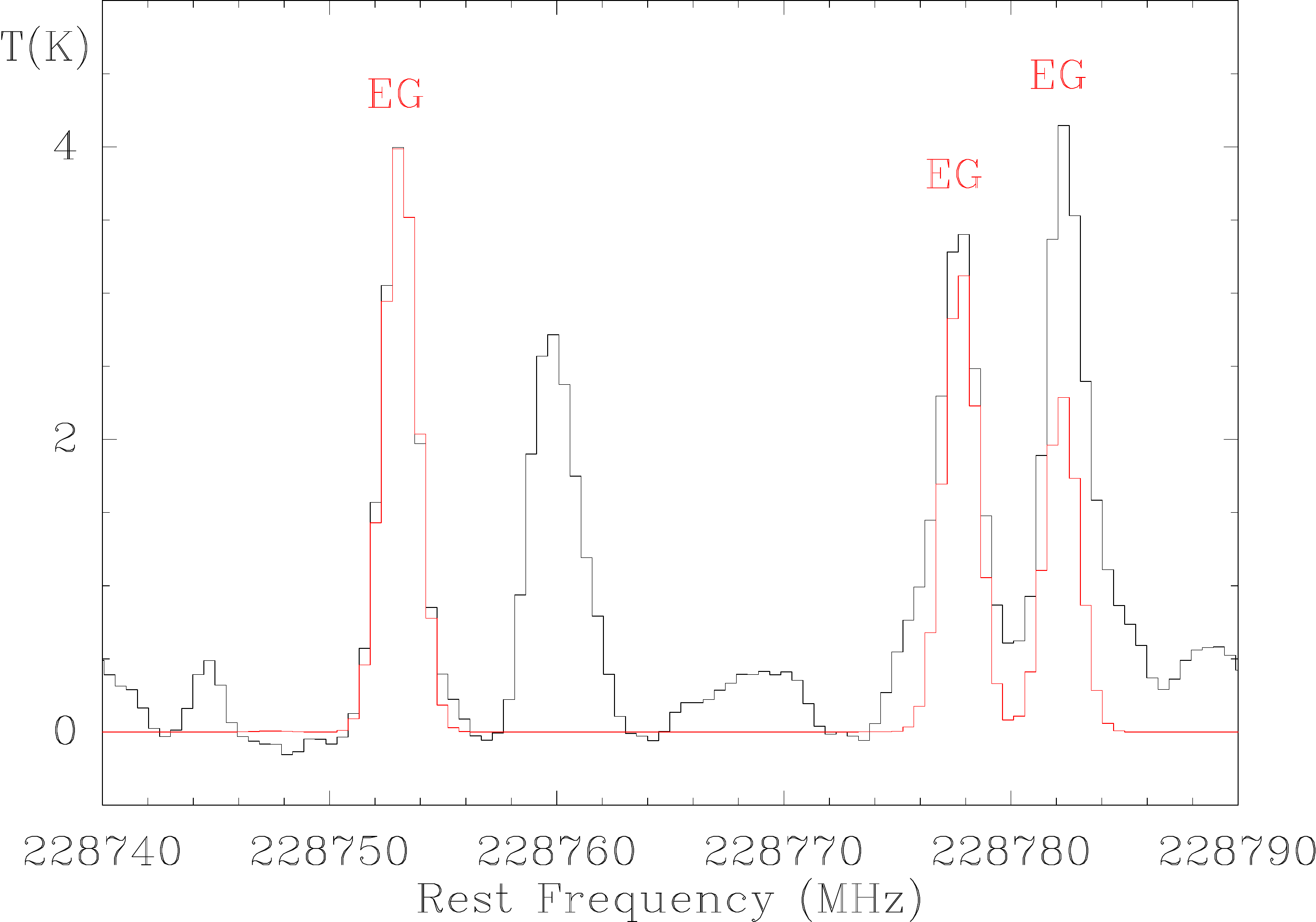}
   \caption{Part of the ALMA spectrum (in black) towards the ethylene glycol peak. The synthetic spectrum of ethylene glycol (EG) is overlaid in red. The parameters used for the synthetic spectrum are: \textit{v}$_\mathrm{LSR}$ = 7.6~km~s$^{-1}$, $\Delta$\textit{v}$_\mathrm{1/2}$ = 2.3~km~s$^{-1}$, \textit{N}$\rm_{\textit{aGg'}}$ = 4.6~10$^{15}~$cm$^{-2}$, \textit{T} = 145~K. Only the line on the left is devoid of confusion.}
         \label{spesyn}
 \end{figure}

\begin{figure}
   \centering
  \includegraphics[width=8cm]{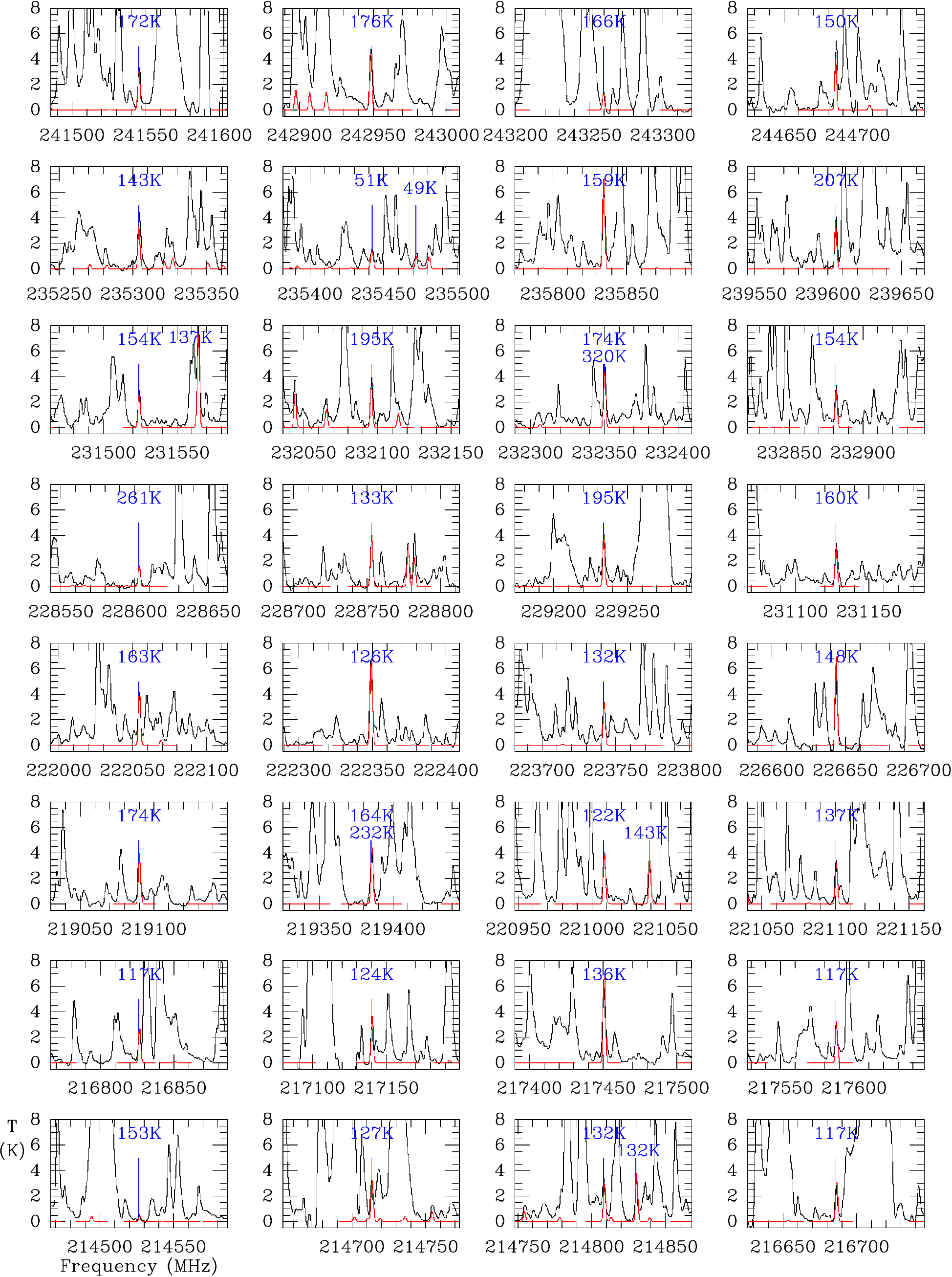}
   \caption{ALMA spectrum (in black) towards the ethylene glycol peak at the ethylene glycol frequencies listed in Table  \ref{table.freq-eg}. The synthetic spectrum of ethylene glycol (EG) is overlaid in red and the value of the upper state energy (E$_{u}$) is indicated in blue.}
         \label{spesyn-table}
 \end{figure}

We calculated the synthetic spectrum for a velocity \textit{v}$_\mathrm{LSR}$ of 7.6~km~s$^{-1}$, a linewidth $\Delta$\textit{v}$_\mathrm{1/2}$ of 2.3~km~s$^{-1}$, a column density \textit{N}$\rm_{\textit{aGg'}}$ of 4.6~10$^{15}~$cm$^{-2}$ and a temperature \textit{T} of 145~K. Figure \ref{spesyn} shows the synthetic spectrum overlaid on the observed ALMA spectrum towards the ethylene glycol peak in the range 228.74--228.79~GHz. Figure \ref{spesyn-table} shows the observed and synthetic spectra of the ethylene glycol transitions listed in Table  \ref{table.freq-eg}. The synthetic spectrum overlaid to the whole ALMA band spectrum is shown in Fig. \ref{spesyn-all}. Figure \ref{spesyn} shows that the transition at 228\,752~MHz is not blended, while the two other lines around 228\,778 and 228\,782~MHz show line contamination and are not included in Table  \ref{table.freq-eg}.

%
\subsection{Search for the \textit{gGg'} conformer of ethylene glycol}
\label{sec:gGg} 

Because the \textit{gGg'} conformer of ethylene glycol is not yet detected in the interstellar medium, we also searched for transitions from this stable conformer although it is 200 cm$^{-1}$ (290~K) higher in energy than the \textit{aGg'} conformer \citep{Muller:2004}. 

All expected strong lines appear to be blended in the ALMA data cube used in this study. Making the simplifying, but admittedly questionable, hypotheses that the \textit{gGg'} conformer presents the same spatial distribution and the same temperature as the \textit{aGg'} conformer, the comparison of the \textit{gGg'} synthetic spectrum with the ALMA observed spectrum at the \textit{aGg'} peak leads to a maximum column density of 10$^{15}~$cm$^{-2}$, i.e. $\le$1/5 of the \textit{aGg'} conformer. We have further assumed here that the partition functions of both conformers are identical since their rotational constants are very similar \citep{Muller:2004}. 

Because the \textit{gGg'} conformer is 290~K higher in energy than \textit{aGg'} and thus may probe hotter gas, we have also derived a synthetic spectrum at the \textit{aGg'} peak for a temperature of 250~K significantly above the 145~K used in the previous Section. We find a projected density upper limit of 2.3~10$^{15}~$cm$^{-2}$. With this higher temperature, we have identified a few lines corresponding to \textit{gGg'} transitions\footnote{Frequencies of \textit{gGg'} conformer transitions corresponding to not very blended lines in the ALMA spectrum at the \textit{aGg'} peak: 220.2498~GHz, 224.8281~GHz, 226.4111~GHz, 227.2706~GHz, 228.9154~GHz, 228.9179~GHz, 228.9374~GHz, 228.9399~GHz, 229.4537~GHz, 229.4540~GHz,230.4622~GHz, 232.7066~GHz, 232.7076~GHz, 232.7080~GHz, 241.6567~GHz.} and with no apparent strong blending. However, the lines are too few and weak to dispel the confusion problem.

The tentative detection of ethylene glycol towards IRAS16293-2422, source B, made by \citet{Jorgensen:2012} corresponds to the 220\,249.8~MHz transition of the \textit{gGg'} conformer. The narrow lines detected towards the IRAS solar-type source B by Jorgensen et al. show little or no line confusion. The presence of ethylene glycol, however, must still await confirmation from detection of other transitions especially those from the most stable \textit{aGg'} conformer. Unfortunately, the 220\,249.8~MHz \textit{gGg'} transition is partially blended in our ALMA spectrum towards \object{Orion-KL} .

\section{Discussion}
\label{sec:discussion} 


\subsection{Comparison with interstellar sources and comets}
\label{sec:comets} 

\begin{table*}[h!]
\begin{minipage}[t]{17cm}
\renewcommand{\footnoterule}{}  
\caption{Comparison of the column densities and abundances of ethylene glycol and related molecules in interstellar sources and comets.}                      
\label{comp-sgrB2}     
\centering        

\begin{tabular}{l c c c c c c}
\hline\hline       
Molecule  &  \multicolumn{3}{c}  {Column density\footnote{See Sect. \ref{sec:comets} for references.}}   & \multicolumn{3}{c}  {Abundance$^a$}\\
& \multicolumn{3}{c}  {(cm$^{-2}$)} & \multicolumn{3}{c} {(\%H$_2$O)}\\
& \object{Orion-KL} & Sgr~B2 & NGC 7129  &  C/1995 O1 &  C/2012 F6  &  C/2013 A1 \\
& & & &  (Hale-Bopp) & (Lemmon) & (Lovejoy)\\
\hline          
(CH$_2$OH)$_2$ &  4.6~10$^{15}$ &  2.3~10$^{15}$ & 2.0~10$^{15}$ & 0.25 & 0.24 & 0.35 \\
CH$_3$OH &  4.2~10$^{18}$ &  5.6~10$^{18}$ & 3.4~10$^{20}$ &  2.4 & 1.6 & 2.6 \\
CH$_3$CH$_2$OH &  2.5~10$^{16}$ &  2.1~10$^{16}$ & 3.0~10$^{18}$ & $\le$0.1 & $\le$0.11 & $\le$0.14 \\
CH$\rm_{2}$OHCHO  & $\le$3.5~10$^{14}$ & 1.8~10$^{15}$ & 1.0~10$^{15}$ & $\le$0.04 & $\le$0.08 & $\le$0.07 \\
\hline      
&  \multicolumn{6}{c}  { Relative abundance with respect to methanol} \\ 
\hline  
(CH$_2$OH)$_2$ &  1.1~10$^{-3}$ &  0.4~10$^{-3}$ & 0.006~10$^{-3}$ & 0.1 & 0.15  & 0.13\\
CH$_3$OH &  1 &   1 & 1 &  1 &  1 &  1 \\
CH$_3$CH$_2$OH &   6~10$^{-3}$&  4~10$^{-3}$ & 9~10$^{-3}$ & $\le$0.04 & $\le$0.07& $\le$0.05\\
CH$\rm_{2}$OHCHO  & $\le$0.08~10$^{-3}$& 0.3~10$^{-3}$& 0.003~10$^{-3}$&  $\le$0.02&  $\le$0.05& $\le$0.03\\

\hline
\end{tabular}
\end{minipage}
\end{table*}


First, we summarize our findings in \object{Orion-KL}. The ethylene glycol peak is close to the prominent methyl formate position MF2 of \citet{Favre:2011a} and the derived temperature for ethylene glycol is similar to that derived for methyl formate. However, the column densities obtained for ethylene glycol, around 4.6 $\times$ 10$^{15}$~cm$^{-2}$, are not very sensitive to the rotational temperature in this temperature range \citep[see Fig. B.6 of][in the case of methanol]{Peng:2012}.
\citet{Favre:2011a} find an upper limit for the glycolaldehyde column density of 3.5 $\times$ 10$^{14}$~cm$^{-2}$ towards MF2 for a temperature of 140~K and a spatial resolution of $3\farcs8\times2\farcs0$ (line confusion does not allow us to get a significant upper limit at a higher spatial resolution). \citet{Peng:2012} derive a column density of $4.2\pm0.4\times10^{18}$~cm$^{-2}$ for the methanol peak named dM-1 (same position as MF2) with the same conditions. Using the ALMA data, we have also studied the ethanol emission (Despois et al. in prep.) and we find a column density of $\sim$ 2.5 $\times$ 10$^{16}$~cm$^{-2}$ and a temperature of $\sim$ 140~K towards the ethanol peak of emission close to the ethylene glycol peak.

Clearly, and as in Sgr~B2 \citep{Belloche:2013} and NGC~7129 FIRS~2 \citep{Fuente:2014}, ethanol is more abundant than ethylene glycol (by a factor of 5 in our data). But ethylene glycol is at least 10 times more abundant than glycolaldehyde in \object{Orion-KL}  whereas it is only twice more abundant in NGC~7129 FIRS~2 and both molecules have similar abundances towards Sgr~B2. In all cases methanol is widespread and highly abundant.

Ethylene glycol is also more abundant than glycolaldehyde in comets and less abundant than methanol \citep[by a factor 7--10,][]{Crovisier:2004a,Biver:2014}, whereas ethanol has not yet been detected with an upper limit 2.5 times lower than ethylene glycol.

Table  \ref{comp-sgrB2} compares the column densities derived for ethylene glycol and related molecules towards \object{Orion-KL}, two other prominent interstellar sources and the 3 comets where ethylene glycol has been identified. We note that the comparison suffers from the fact that the spatial distribution is different for the different molecules and the spatial resolution is also different between the observed interstellar sources.

With the recent detection of ethylene glycol in two comets \citep{Biver:2014}, whereas more abundant interstellar species like ethanol and glycolaldehyde are still undetected, the problem of ethylene glycol formation gets even more acute. Pathways to its formation are still under investigation. \citet{Walsh:2014} summarize available grain-surface scenarios which include a) formation from CH$_{2}$OH radical, b) hydrogenation of OCCHO itself resulting from atomic oxygen addition on H$_{2}$CO, c) some non detailed UV or CR processing of ices, d) possibility of thermal processing.

\subsection{\object{Orion-KL} }
\label{sec:orion} 

Comparison of the ethylene glycol distribution in \object{Orion-KL}  with other available physical or chemical tracers may shed some light on the molecular formation conditions in \object{Orion-KL}. We compared the maps of these tracers using the Aladin software\footnote{Centre de Donn\'ees Astronomiques de Strasbourg, http://aladin.u-strasbg.fr/aladin.gml}    
\citep{Bonnarel:2000}. This in turn could help us to understand the molecular complexity in other regions and perhaps in comets. 

There is no obvious role played by temperature: the ethylene glycol peak does not show up in any of the high-resolution temperature maps of CH$_{3}$OH \citep{Beuther:2006,Friedel:2012} and NH$_{3}$ \citep{Goddi:2011b} or in any Mid-IR peaks \citep{Okumura:2011}.

The various tracers do not show any concentration of matter at the ethylene glycol peak. On the contrary, high-resolution maps of $^{13}$CO \citep{Zapata:2010a} and continuum emission present a hole at the ethylene glycol peak and no clear association is seen with NH$_{3}$ \citep{Wilson:2000}.

At higher spatial resolution two continuum spots, C21 and C22 \citep{Friedel:2011}, appear on each side of the ethylene glycol peak. The northern spot is close to the SMA1 continuum source and coincides with a 22~GHz H$_{2}$O maser \citep{Gaume:1998} and the MF6 methyl formate position of \citet{Favre:2011a}. There is another 22~GHz maser spot in the south at the same position as the MF2 methyl formate peak near but not exactly coincident with the continuum peak.

In the near-IR continuum as well as in H$_{2}$ 2.2~$\mu$m \citep[][D. Rouan, private communication]{Lacombe:2004}, the region appears as a dark patch; this is also the case at longer wavelength  \citep[e.g. at 20~$\mu$m,][]{Robberto:2005}.

One short CO jet among jets identified by \citet{Zapata:2009} (and coming from the common "explosion centre" in between source I and BN)  overlaps the ethylene glycol peak.

A precise comparison with ALMA maps of other oxygenated species is in preparation. According to previous studies by us and other groups, it appears clearly that the ethylene glycol distribution is definitely different from that of methyl formate and dimethyl ether \citep[see e.g.][]{Brouillet:2013}. It is also different from CN bearing species, such as C$_{2}$H$_{5}$CN, from acetone CH$_{3}$COCH$_{3}$ \citep{Peng:2013,Friedel:2008,Widicus-Weaver:2012}, and from deuterated methanol \citep{Peng:2012}. One of the $^{13}$C methanol peaks in \citet{Peng:2012} is however coincident with the ethylene glycol peak. Compared with HDO, which is typically released from grain surfaces, ethylene glycol appears to come from the same region as the main HDO emission delineated in the maps of \citet{Neill:2013} obtained from the ALMA-SV data. However, looking at the individual channel maps of the two molecules, we find at each velocity that the ethylene glycol emission peak is $\sim$1$\arcsec$ west of the HDO peak.

These first comparisons of maps: a) seem to exclude that the temperature deduced from present-day observations could play a dominant role in the observed differences  (but an earlier gas warming at ethylene glycol position is not excluded); and b) show that ethylene glycol is distinct from species like dimethyl ether and methyl formate, which are strong in the compact ridge and likely to be produced on grain mantles. The compact nature of the ethylene glycol emission compared to the more extended distributions of methanol and ethanol calls for a special production scenario. The above elements suggest that one possibility could be a collision of a high-velocity jet identified in CO (but possibly ionized) with a clump at the edge or in front of the main hot core condensation, which would be responsible for the ethylene glycol production and the hole in continuum emission.  This collision could also be favourable to the excitation of the two 22~GHz maser spots. Another scenario, involving UV photons and reactions with radicals, is not excluded. However,  there is no special observational evidence in favour of this scenario, as no local source emitting strong UV emission has been identified yet in this area, and penetration of external UV photons into such a dense region would require voids in the dust distribution. The association of two CH$_{2}$OH radicals to form (CH$_{2}$OH)$_{2}$, although appealing by its simplicity, is not favoured as the CH$_{2}$OH radical has a low mobility due to OH bonds \citep{Walsh:2014}.
\section{Continuum emission}
\label{sec:continuum}

\begin{figure*}
   \centering
    \includegraphics[width=6cm]{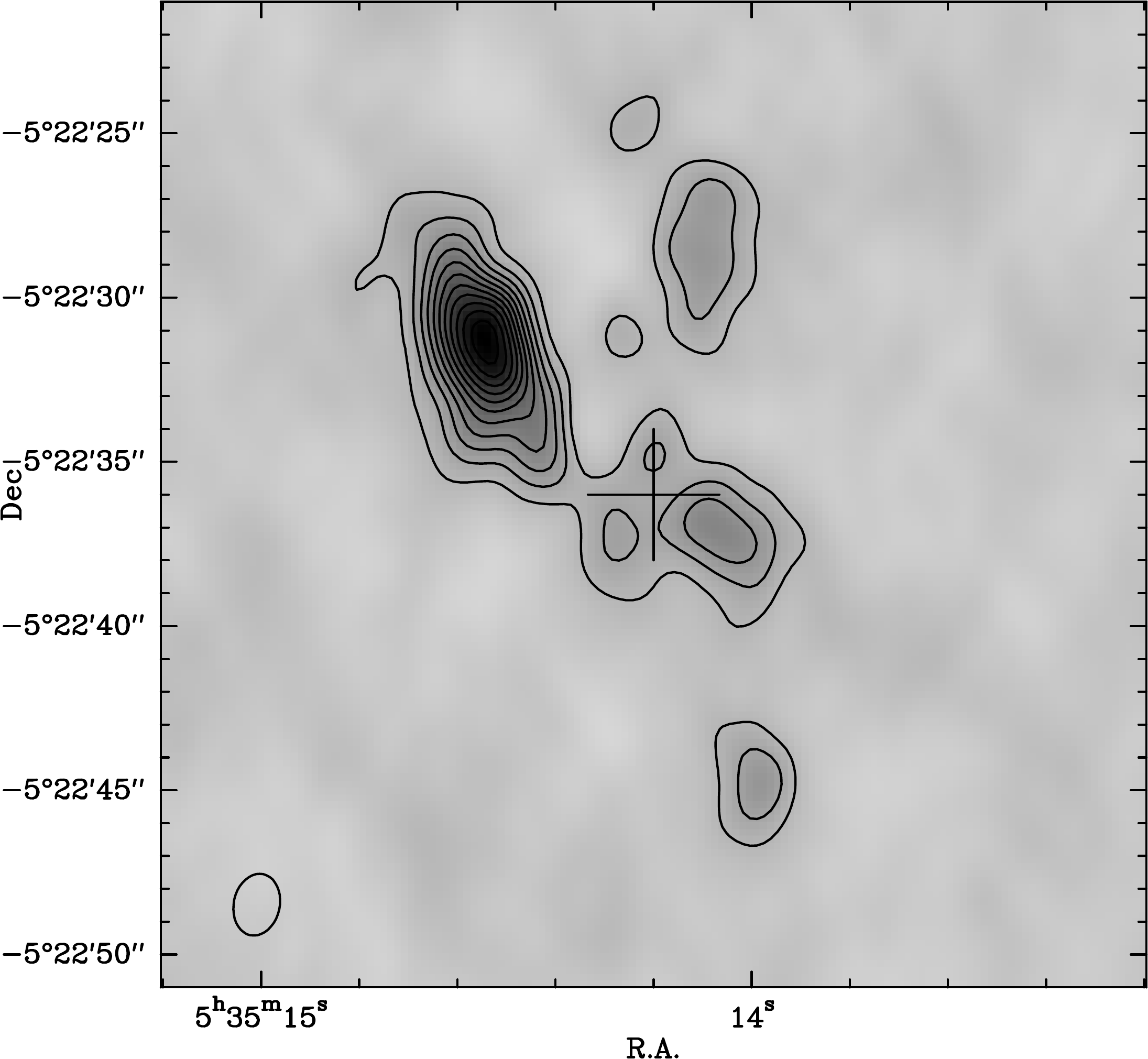}  
  \includegraphics[width=6cm]{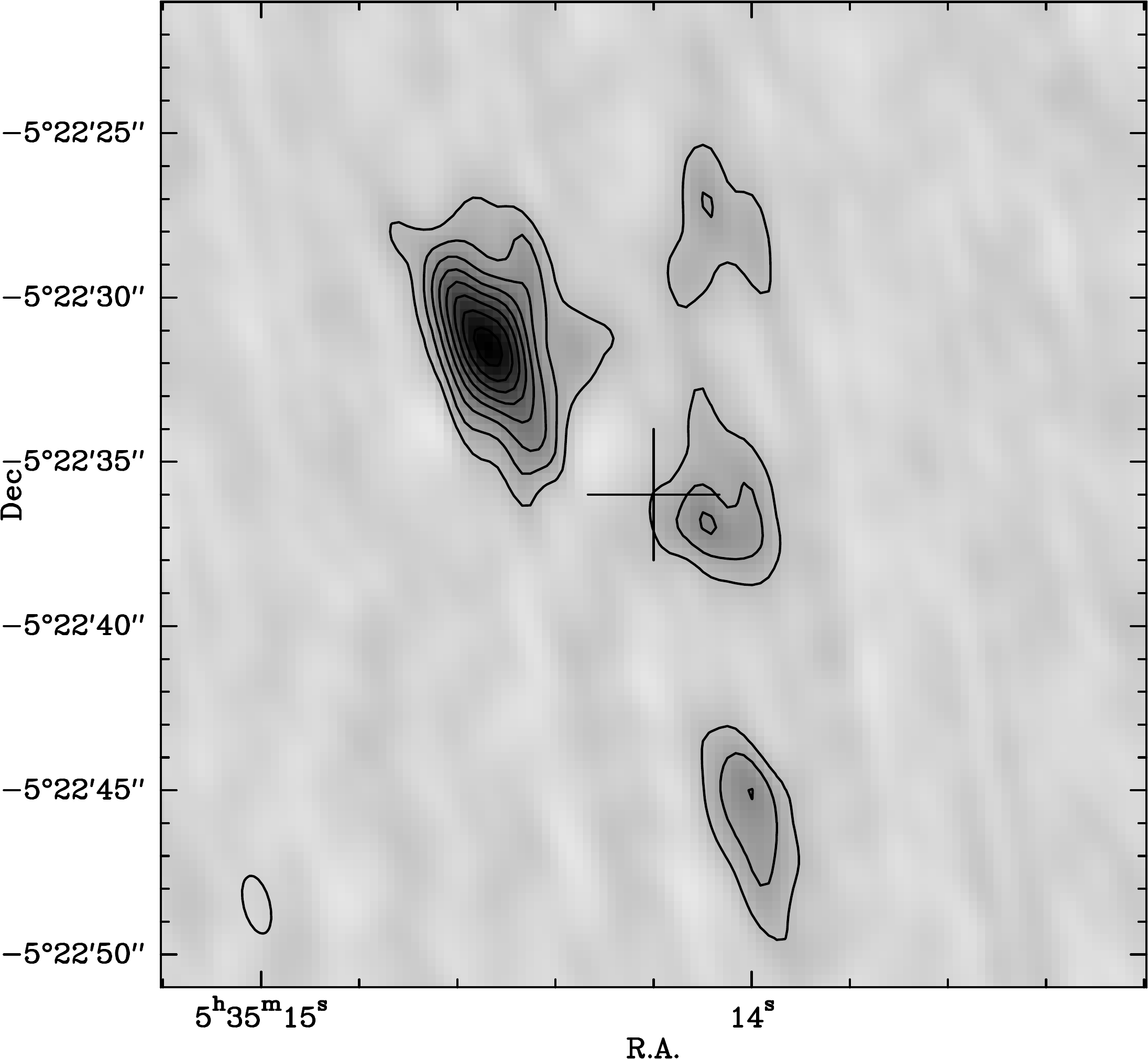}

   \caption{Left: continuum map obtained with ALMA at 223\,GHz. The beam shown in the bottom left corner is 1.92$\arcsec$ $\times$ 1.38$\arcsec$. Right: continuum map obtained with the IRAM Plateau de Bure Interferometer at 223\,GHz \citep{Favre:2011a}. The beam shown in the bottom left corner is 1.79$\arcsec$ $\times$ 0.79$\arcsec$. The level step and first contour for the two maps are 1\,K.}
         \label{continuum-maps}
 \end{figure*}

A map of the continuum emission at 1.3\,mm in \object{Orion-KL}  is a by-product of our efforts to isolate the line emission by subtracting the underlying continuum. The ALMA data covers a total bandwidth of 32.9~GHz centred at 230.171~GHz. For practical reasons, we have separated each one of the 20 spectral windows into four 1\,000-channel sub-windows, and we have searched for line-free channels in each sub-window using the following method.

If a channel map has no noticeable line emission, the difference between the maximum and the minimum intensities of the channel map will be lower than for channels with stronger molecular emission. Sorting out channels by increasing value of this difference, we have retained the best channel candidates to be line-free.

We then used the fact that if two channel maps are devoid (within the noise) of molecular emission, their difference is noise plus the difference in continuum emission. This contribution is negligible if the channel frequencies are close enough (as is the case in a sub-window). We selected the best of channel pairs with a low and spatially uniform noise difference.

Next, we eliminated channel pairs whose frequencies are closer than a typical line width, and might thus have identical molecular emission, inappropriately passing our second criterion.

Then we obtained the continuum map for a given sub-window by averaging the two channel maps of the best pair. 

Finally, we checked the coherence of the continuum maps obtained for adjacent sub-windows.

It was not possible to identify line-free channels for all the sub-windows, in that case, we took the continuum emission from a sub-window adjacent in frequency. We have thus obtained  an 80-channel continuum cube in the range 213.7~GHz to 246.6~GHz. 

Figure \ref{continuum-maps} compares the continuum maps obtained at the same frequency and with a similar spatial resolution with ALMA (this work) and the IRAM Plateau de Bure Interferometer \citep{Favre:2011a}. There is a good agreement between both maps which show the same four main components. The ALMA Hot Core position is ($\alpha_{J2000}$ = 05$^{h}$35$^{m}$14$\fs$55, $\delta_{J2000}$ = $-$05$\degr$22$\arcmin$31$\farcs$3) and it corresponds, within 0.2$\arcsec$, to the Ca continuum peak identified in \citet{Favre:2011a}. The slight differences observed in the ALMA and IRAM interferometer maps are likely due to different spatial filtering of the two arrays. 

\begin{figure}
   \centering
  \includegraphics[width=4.2cm]{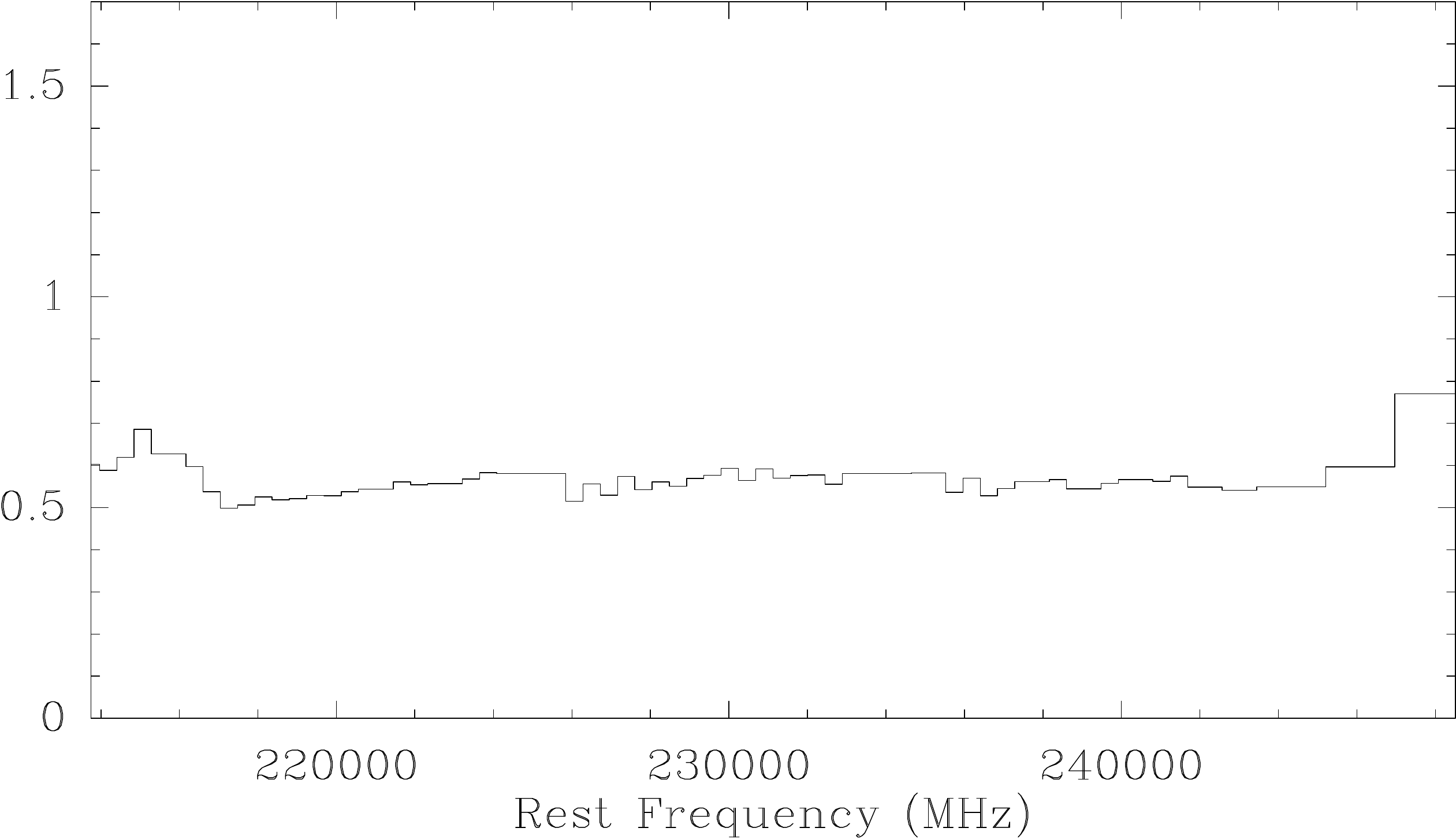}  
  \includegraphics[width=4.2cm]{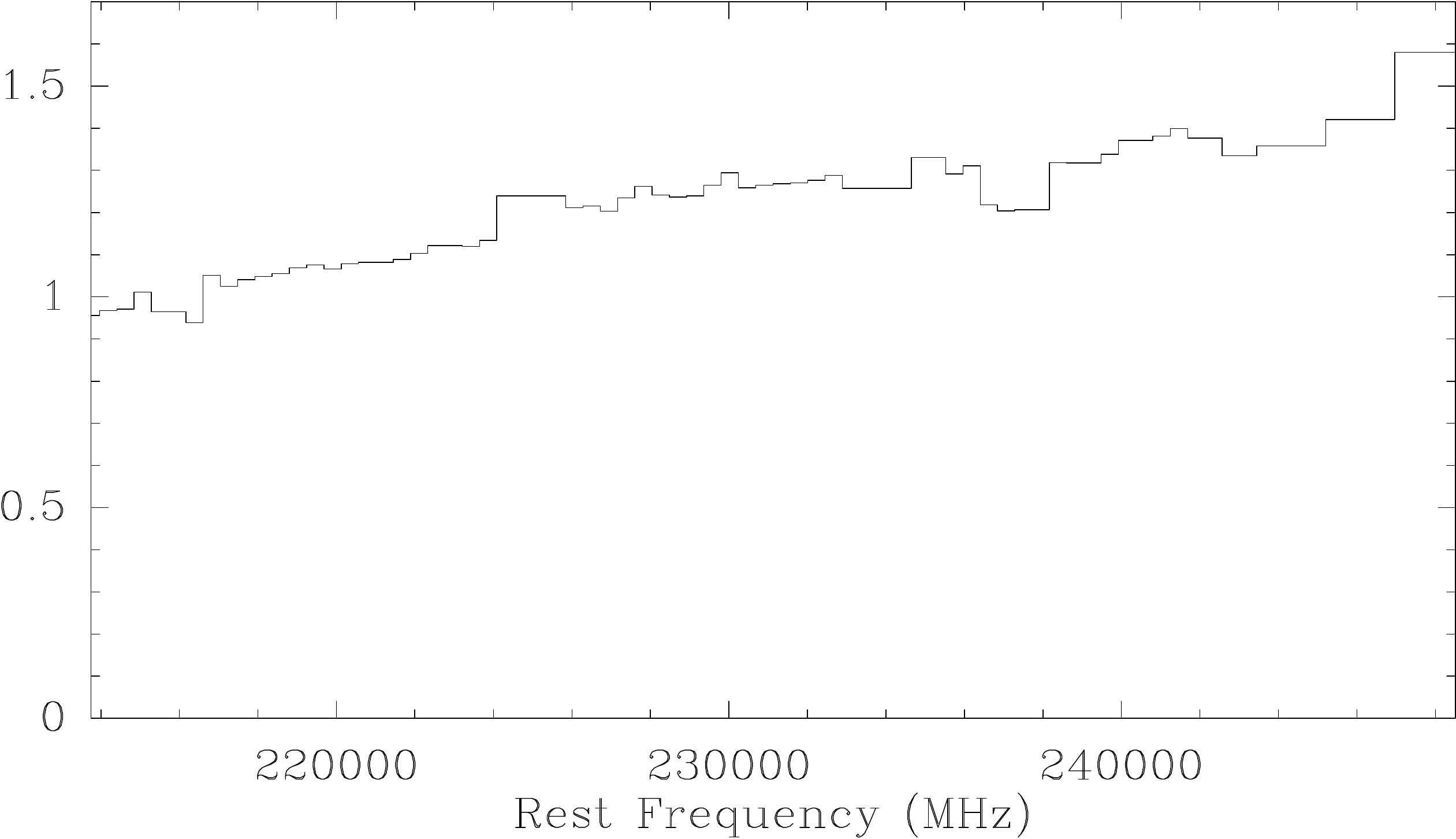}
   \caption{Flux density per beam (in Jy/beam) versus frequency across ALMA band 6. Left: at the position of the ethylene glycol peak. Right: at the Hot Core position. }
         \label{continuum-spectra}
 \end{figure}

Figure \ref{continuum-spectra} shows the continuum variation with respect to the frequency over the whole ALMA-SV bandwidth. The spectrum is flat at the position of the ethylene glycol peak whereas it rises with frequency at the Hot Core position. 
This seems to be related to the sizes of the different continuum sub-sources associated with our main continuum component. In particular, if the continuum sub-source close to the Hot Core position is more compact than the synthesized beam, its intensity will vary as the beam size varies across the frequency range covered by the ALMA-SV data. We further note that our unresolved main continuum component (Hot Core) shows a NE-SW extension
which encompasses the two sub-sources seen in the sub-arcsec 870~$\mu$m dust continuum map of \citet{Tang:2010} and the complex emission recently mapped with 0.5$\arcsec$ ALMA resolution by \citet{Hirota:2015}.

\section{Conclusions}
\label{sec:conclusion}
  \begin{enumerate}
      \item We detected the \textit{aGg'} conformer of ethylene glycol in Orion~KL. The emission peaks towards the Hot Core close to the main continuum peak, about 2$\arcsec$ to the south-west. The distribution is compact with a deconvolved size $\le$ 2.4$\arcsec$$\times$ 1.1$\arcsec$. It is clearly different from that of other complex O-bearing species like methyl formate, dimethyl ether, acetone, and ethanol.
      \item We deduced a beam averaged column density \textit{N}$\rm_{\textit{aGg'}}$ of 4.6$\pm$0.8~10$^{15}~$cm$^{-2}$ and a rotational temperature \textit{T} of 145$\pm$30~K from a population diagram.  Using the continuum map, we find an ethylene glycol abundance with respect to H$_{2}$ of the order of 10$^{-9}$.
      \item  Taking methanol CH$_{3}$OH as a reference, the relative abundance of ethylene glycol is of the order of 10$^{-3}$  in \object{Orion-KL}, similar to that in Sgr~B2. We confirm with \object{Orion-KL} that the ethylene glycol abundance in interstellar sources is much lower than in the three comets where it has been detected with a ratio in the range 0.1--0.15. This result places ethylene glycol clearly outside the correlation found by \citet{Bockelee-Morvan:2000} between Comet C/1995 O1 (Hale-Bopp) and four well-studied compact sources, including \object{Orion-KL}.
	 \item The simple alcohol ethanol CH$_{3}$CH$_{2}$OH  is more abundant than the di-alcohol CH$_{2}$OHCH$_{2}$OH  by a factor of 5 in \object{Orion-KL} and even larger factors were found in Sgr~B2 and
NGC~7129. This result confirms a marked difference with the comets where, on the opposite, an upper limit of 0.5 has
been derived for the ethanol/ethylene glycol ratio. 
      \item We only identified weak lines of the \textit{gGg'} conformer at the expected frequencies, and all strong lines were blended. Hence we conclude that the \textit{gGg'} conformer is not detected in our study, and we derive a column density less than 0.2 times the \textit{aGg'} conformer's at the emission peak of the latter.
      \item The identification of the ethylene glycol \textit{aGg'} conformer was made possible thanks to the ALMA-SV data. A large frequency range is necessary to identify complex molecules and interferometric maps of the transitions considerably helped us in diminishing the line confusion problem.     
      \item Molecules have different spatial distribution even when a priori of a same group (e.g. large O-bearing molecules). This should be taken into account in future comparisons between interstellar sources observed at different spatial resolution as well as between interstellar sources and comets. It might bring diversity in forming planets and their content in prebiotic species. 

   \end{enumerate}

\begin{acknowledgements}
      We thank the referee for his help in improving the paper. This paper makes use of the following ALMA data: ADS/JAO.ALMA\#2011.0.00009.SV. ALMA is a partnership of ESO (representing its member states), NSF (USA) and NINS (Japan), together with NRC (Canada) and NSC and ASIAA (Taiwan), in cooperation with the Republic of Chile. The Joint ALMA Observatory is operated by ESO, AUI/NRAO and NAOJ. We thank the IRAM staff in Grenoble and, in particular, Jan-Martin Winter. This study started with IRAM PdBI data kindly provided by T. Jacq. M. Ali Dib's internship was of great help in the first developments and tests of the continuum subtraction method. This work was supported by CNRS national programs PCMI (Physics and Chemistry of the Interstellar Medium) and GDR Exobiology. 
      \end{acknowledgements}

%
%
\bibliographystyle{aa}
\bibliography{biblio}

\begin{thebibliography}{54}
\expandafter\ifx\csname natexlab\endcsname\relax\def\natexlab#1{#1}\fi

\bibitem[{{Belloche} {et~al.}(2013){Belloche}, {M{\"u}ller}, {Menten},
  {Schilke}, \& {Comito}}]{Belloche:2013}
{Belloche}, A., {M{\"u}ller}, H.~S.~P., {Menten}, K.~M., {Schilke}, P., \&
  {Comito}, C. 2013, \aap, 559, A47

\bibitem[{{Beuther} {et~al.}(2006){Beuther}, {Zhang}, {Reid}, {Hunter},
  {Gurwell}, {Wilner}, {Zhao}, {Shinnaga}, {Keto}, {Ho}, {Moran}, \&
  {Liu}}]{Beuther:2006}
{Beuther}, H., {Zhang}, Q., {Reid}, M.~J., {et~al.} 2006, \apj, 636, 323

\bibitem[{{Biver} {et~al.}(2014){Biver}, {Bockel{\'e}e-Morvan}, {Debout},
  {Crovisier}, {Boissier}, {Lis}, {Dello Russo}, {Moreno}, {Colom}, {Paubert},
  {Vervack}, \& {Weaver}}]{Biver:2014}
{Biver}, N., {Bockel{\'e}e-Morvan}, D., {Debout}, V., {et~al.} 2014, \aap, 566,
  L5

\bibitem[{{Bockel{\'e}e-Morvan} {et~al.}(2002){Bockel{\'e}e-Morvan}, {Gautier},
  {Hersant}, {Hur{\'e}}, \& {Robert}}]{Bockelee-Morvan:2002}
{Bockel{\'e}e-Morvan}, D., {Gautier}, D., {Hersant}, F., {Hur{\'e}}, J.-M., \&
  {Robert}, F. 2002, \aap, 384, 1107

\bibitem[{{Bockel{\'e}e-Morvan} {et~al.}(2000){Bockel{\'e}e-Morvan}, {Lis},
  {Wink}, {Despois}, {Crovisier}, {Bachiller}, {Benford}, {Biver}, {Colom},
  {Davies}, {G{\'e}rard}, {Germain}, {Houde}, {Mehringer}, {Moreno}, {Paubert},
  {Phillips}, \& {Rauer}}]{Bockelee-Morvan:2000}
{Bockel{\'e}e-Morvan}, D., {Lis}, D.~C., {Wink}, J.~E., {et~al.} 2000, \aap,
  353, 1101

\bibitem[{{Bonnarel} {et~al.}(2000){Bonnarel}, {Fernique}, {Bienaym{\'e}},
  {Egret}, {Genova}, {Louys}, {Ochsenbein}, {Wenger}, \&
  {Bartlett}}]{Bonnarel:2000}
{Bonnarel}, F., {Fernique}, P., {Bienaym{\'e}}, O., {et~al.} 2000, \aaps, 143,
  33

\bibitem[{{Brouillet} {et~al.}(2013){Brouillet}, {Despois}, {Baudry}, {Peng},
  {Favre}, {Wootten}, {Remijan}, {Wilson}, {Combes}, \&
  {Wlodarczak}}]{Brouillet:2013}
{Brouillet}, N., {Despois}, D., {Baudry}, A., {et~al.} 2013, \aap, 550, A46

\bibitem[{{Ceccarelli} {et~al.}(2014){Ceccarelli}, {Caselli},
  {Bockelee-Morvan}, {Mousis}, {Pizzarello}, {Robert}, \&
  {Semenov}}]{Ceccarelli:2014}
{Ceccarelli}, C., {Caselli}, P., {Bockelee-Morvan}, D., {et~al.} 2014,
  Protostars and Planets VI, in press (arXiv:1403.7143)

\bibitem[{{Christen} {et~al.}(2001){Christen}, {Coudert}, {Larsson}, \&
  {Cremer}}]{Christen:2001}
{Christen}, D., {Coudert}, L.~H., {Larsson}, J.~A., \& {Cremer}, D. 2001,
  Journal of Molecular Spectroscopy, 205, 185

\bibitem[{{Christen} {et~al.}(1995){Christen}, {Coudert}, {Suenram}, \&
  {Lovas}}]{Christen:1995}
{Christen}, D., {Coudert}, L.~H., {Suenram}, R.~D., \& {Lovas}, F.~J. 1995,
  Journal of Molecular Spectroscopy, 172, 57

\bibitem[{{Christen} \& {M{\"u}ller}(2003)}]{Christen:2003}
{Christen}, D. \& {M{\"u}ller}, H.~S.~P. 2003, Physical Chemistry Chemical
  Physics (Incorporating Faraday Transactions), 5, 3600

\bibitem[{{Clark}(1980)}]{Clark:1980}
{Clark}, B.~G. 1980, \aap, 89, 377

\bibitem[{{Crovisier} {et~al.}(2004{\natexlab{a}}){Crovisier},
  {Bockel{\'e}e-Morvan}, {Biver}, {Colom}, {Despois}, \&
  {Lis}}]{Crovisier:2004}
{Crovisier}, J., {Bockel{\'e}e-Morvan}, D., {Biver}, N., {et~al.}
  2004{\natexlab{a}}, \aap, 418, L35

\bibitem[{{Crovisier} {et~al.}(2004{\natexlab{b}}){Crovisier},
  {Bockel{\'e}e-Morvan}, {Colom}, {Biver}, {Despois}, {Lis}, \& {the Team for
  target-of-opportunity radio observations of comets}}]{Crovisier:2004a}
{Crovisier}, J., {Bockel{\'e}e-Morvan}, D., {Colom}, P., {et~al.}
  2004{\natexlab{b}}, \aap, 418, 1141

\bibitem[{{Crovisier} {et~al.}(1997){Crovisier}, {Leech}, {Bockelee-Morvan},
  {Brooke}, {Hanner}, {Altieri}, {Keller}, \& {Lellouch}}]{Crovisier:1997}
{Crovisier}, J., {Leech}, K., {Bockelee-Morvan}, D., {et~al.} 1997, Science,
  275, 1904

\bibitem[{{Favre} {et~al.}(2011){Favre}, {Despois}, {Brouillet}, {Baudry},
  {Combes}, {Gu\'elin}, {Wootten}, \& {Wlodarczak}}]{Favre:2011a}
{Favre}, C., {Despois}, D., {Brouillet}, N., {et~al.} 2011, \aap, 532, 32

\bibitem[{{Friedel} \& {Snyder}(2008)}]{Friedel:2008}
{Friedel}, D.~N. \& {Snyder}, L.~E. 2008, \apj, 672, 962

\bibitem[{{Friedel} {et~al.}(2004){Friedel}, {Snyder}, {Turner}, \&
  {Remijan}}]{Friedel:2004}
{Friedel}, D.~N., {Snyder}, L.~E., {Turner}, B.~E., \& {Remijan}, A. 2004,
  \apj, 600, 234

\bibitem[{{Friedel} \& {Widicus Weaver}(2011)}]{Friedel:2011}
{Friedel}, D.~N. \& {Widicus Weaver}, S.~L. 2011, \apj, 742, 64

\bibitem[{{Friedel} \& {Widicus Weaver}(2012)}]{Friedel:2012}
{Friedel}, D.~N. \& {Widicus Weaver}, S.~L. 2012, \apjs, 201, 17

\bibitem[{{Fuente} {et~al.}(2014){Fuente}, {Cernicharo}, {Caselli}, {McCoey},
  {Johnstone}, {Fich}, {van Kempen}, {Palau}, {Y{\i}ld{\i}z}, {Tercero}, \&
  {L{\'o}pez}}]{Fuente:2014}
{Fuente}, A., {Cernicharo}, J., {Caselli}, P., {et~al.} 2014, \aap, 568, A65

\bibitem[{{Gaume} {et~al.}(1998){Gaume}, {Wilson}, {Vrba}, {Johnston}, \&
  {Schmid-Burgk}}]{Gaume:1998}
{Gaume}, R.~A., {Wilson}, T.~L., {Vrba}, F.~J., {Johnston}, K.~J., \&
  {Schmid-Burgk}, J. 1998, \apj, 493, 940

\bibitem[{{Goddi} {et~al.}(2011{\natexlab{a}}){Goddi}, {Greenhill},
  {Humphreys}, {Chandler}, \& {Matthews}}]{Goddi:2011b}
{Goddi}, C., {Greenhill}, L.~J., {Humphreys}, E.~M.~L., {Chandler}, C.~J., \&
  {Matthews}, L.~D. 2011{\natexlab{a}}, \apjl, 739, L13

\bibitem[{{Goddi} {et~al.}(2011{\natexlab{b}}){Goddi}, {Humphreys},
  {Greenhill}, {Chandler}, \& {Matthews}}]{Goddi:2011a}
{Goddi}, C., {Humphreys}, E.~M.~L., {Greenhill}, L.~J., {Chandler}, C.~J., \&
  {Matthews}, L.~D. 2011{\natexlab{b}}, \apj, 728, 15

\bibitem[{{Goldsmith} \& {Langer}(1999)}]{Goldsmith:1999}
{Goldsmith}, P.~F. \& {Langer}, W.~D. 1999, \apj, 517, 209

\bibitem[{{Gu{\'e}lin} {et~al.}(2008){Gu{\'e}lin}, {Brouillet}, {Cernicharo},
  {Combes}, \& {Wooten}}]{Guelin:2008}
{Gu{\'e}lin}, M., {Brouillet}, N., {Cernicharo}, J., {Combes}, F., \& {Wooten},
  A. 2008, \apss, 313, 45

\bibitem[{{Halfen} {et~al.}(2006){Halfen}, {Apponi}, {Woolf}, {Polt}, \&
  {Ziurys}}]{Halfen:2006}
{Halfen}, D.~T., {Apponi}, A.~J., {Woolf}, N., {Polt}, R., \& {Ziurys}, L.~M.
  2006, \apj, 639, 237

\bibitem[{{Herbst} \& {van Dishoeck}(2009)}]{Herbst:2009}
{Herbst}, E. \& {van Dishoeck}, E.~F. 2009, \araa, 47, 427

\bibitem[{{Hirota} {et~al.}(2015){Hirota}, {Kim}, {Kurono}, \&
  {Honma}}]{Hirota:2015}
{Hirota}, T., {Kim}, M.~K., {Kurono}, Y., \& {Honma}, M. 2015, \apj, in press

\bibitem[{{Hollis} {et~al.}(2002){Hollis}, {Lovas}, {Jewell}, \&
  {Coudert}}]{Hollis:2002}
{Hollis}, J.~M., {Lovas}, F.~J., {Jewell}, P.~R., \& {Coudert}, L.~H. 2002,
  \apjl, 571, L59

\bibitem[{{Irvine} {et~al.}(2000){Irvine}, {Schloerb}, {Crovisier}, {Fegley},
  \& {Mumma}}]{Irvine:2000}
{Irvine}, W.~M., {Schloerb}, F.~P., {Crovisier}, J., {Fegley}, Jr., B., \&
  {Mumma}, M.~J. 2000, Protostars and Planets IV, 1159

\bibitem[{{J{\o}rgensen} {et~al.}(2012){J{\o}rgensen}, {Favre}, {Bisschop},
  {Bourke}, {van Dishoeck}, \& {Schmalzl}}]{Jorgensen:2012}
{J{\o}rgensen}, J.~K., {Favre}, C., {Bisschop}, S.~E., {et~al.} 2012, \apjl,
  757, L4

\bibitem[{{Lacombe} {et~al.}(2004){Lacombe}, {Gendron}, {Rouan}, {Cl{\'e}net},
  {Field}, {Lemaire}, {Gustafsson}, {Lagrange}, {Mouillet}, {Rousset}, {Fusco},
  {Rousset-Rouvi{\`e}re}, {Servan}, {Marlot}, \& {Feautrier}}]{Lacombe:2004}
{Lacombe}, F., {Gendron}, E., {Rouan}, D., {et~al.} 2004, \aap, 417, L5

\bibitem[{{Maury} {et~al.}(2014){Maury}, {Belloche}, {Andr{\'e}}, {Maret},
  {Gueth}, {Codella}, {Cabrit}, {Testi}, \& {Bontemps}}]{Maury:2014}
{Maury}, A.~J., {Belloche}, A., {Andr{\'e}}, P., {et~al.} 2014, \aap, 563, L2

\bibitem[{{M{\"u}ller} \& {Christen}(2004)}]{Muller:2004}
{M{\"u}ller}, H.~S.~P. \& {Christen}, D. 2004, Journal of Molecular
  Spectroscopy, 228, 298

\bibitem[{{M{\"u}ller} {et~al.}(2005){M{\"u}ller}, {Schl{\"o}der}, {Stutzki},
  \& {Winnewisser}}]{Muller:2005}
{M{\"u}ller}, H.~S.~P., {Schl{\"o}der}, F., {Stutzki}, J., \& {Winnewisser}, G.
  2005, Journal of Molecular Structure, 742, 215

\bibitem[{{M{\"u}ller} {et~al.}(2001){M{\"u}ller}, {Thorwirth}, {Roth}, \&
  {Winnewisser}}]{Muller:2001}
{M{\"u}ller}, H.~S.~P., {Thorwirth}, S., {Roth}, D.~A., \& {Winnewisser}, G.
  2001, \aap, 370, L49

\bibitem[{{Mumma} \& {Charnley}(2011)}]{Mumma:2011}
{Mumma}, M.~J. \& {Charnley}, S.~B. 2011, \araa, 49, 471

\bibitem[{{Neill} {et~al.}(2013){Neill}, {Wang}, {Bergin}, {Crockett}, {Favre},
  {Plume}, \& {Melnick}}]{Neill:2013}
{Neill}, J.~L., {Wang}, S., {Bergin}, E.~A., {et~al.} 2013, \apj, 770, 142

\bibitem[{{Nummelin} {et~al.}(2000){Nummelin}, {Bergman}, {Hjalmarson},
  {Friberg}, {Irvine}, {Millar}, {Ohishi}, \& {Saito}}]{Nummelin:2000}
{Nummelin}, A., {Bergman}, P., {Hjalmarson}, {\AA}., {et~al.} 2000, \apjs, 128,
  213

\bibitem[{{Okumura} {et~al.}(2011){Okumura}, {Yamashita}, {Sako}, {Miyata},
  {Honda}, {Kataza}, \& {Okamoto}}]{Okumura:2011}
{Okumura}, S.-I., {Yamashita}, T., {Sako}, S., {et~al.} 2011, \pasj, 63, 823

\bibitem[{{Peng} {et~al.}(2013){Peng}, {Despois}, {Brouillet}, {Baudry},
  {Favre}, {Remijan}, {Wootten}, {Wilson}, {Combes}, \&
  {Wlodarczak}}]{Peng:2013}
{Peng}, T.-C., {Despois}, D., {Brouillet}, N., {et~al.} 2013, \aap, 554, A78

\bibitem[{{Peng} {et~al.}(2012){Peng}, {Despois}, {Brouillet}, {Parise}, \&
  {Baudry}}]{Peng:2012}
{Peng}, T.-C., {Despois}, D., {Brouillet}, N., {Parise}, B., \& {Baudry}, A.
  2012, \aap, 543, A152

\bibitem[{{Requena-Torres} {et~al.}(2008){Requena-Torres},
  {Mart{\'{\i}}n-Pintado}, {Mart{\'{\i}}n}, \& {Morris}}]{requena-torres:2008}
{Requena-Torres}, M.~A., {Mart{\'{\i}}n-Pintado}, J., {Mart{\'{\i}}n}, S., \&
  {Morris}, M.~R. 2008, \apj, 672, 352

\bibitem[{{Robberto} {et~al.}(2005){Robberto}, {Beckwith}, {Panagia}, {Patel},
  {Herbst}, {Ligori}, {Custo}, {Boccacci}, \& {Bertero}}]{Robberto:2005}
{Robberto}, M., {Beckwith}, S.~V.~W., {Panagia}, N., {et~al.} 2005, \aj, 129,
  1534

\bibitem[{{Snyder} {et~al.}(2005){Snyder}, {Lovas}, {Hollis}, {Friedel},
  {Jewell}, {Remijan}, {Ilyushin}, {Alekseev}, \& {Dyubko}}]{Snyder:2005}
{Snyder}, L.~E., {Lovas}, F.~J., {Hollis}, J.~M., {et~al.} 2005, \apj, 619, 914

\bibitem[{{Tang} {et~al.}(2010){Tang}, {Ho}, {Koch}, \& {Rao}}]{Tang:2010}
{Tang}, Y., {Ho}, P.~T.~P., {Koch}, P.~M., \& {Rao}, R. 2010, \apj, 717, 1262

\bibitem[{{Tercero} {et~al.}(2010){Tercero}, {Cernicharo}, {Pardo}, \&
  {Goicoechea}}]{Tercero:2010}
{Tercero}, B., {Cernicharo}, J., {Pardo}, J.~R., \& {Goicoechea}, J.~R. 2010,
  \aap, 517, A96

\bibitem[{{Walsh} {et~al.}(2014){Walsh}, {Millar}, {Nomura}, {Herbst}, {Widicus
  Weaver}, {Aikawa}, {Laas}, \& {Vasyunin}}]{Walsh:2014}
{Walsh}, C., {Millar}, T.~J., {Nomura}, H., {et~al.} 2014, \aap, 563, A33

\bibitem[{{Widicus Weaver} \& {Friedel}(2012)}]{Widicus-Weaver:2012}
{Widicus Weaver}, S.~L. \& {Friedel}, D.~N. 2012, \apjs, 201, 16

\bibitem[{{Wilson} {et~al.}(2000){Wilson}, {Gaume}, {Gensheimer}, \&
  {Johnston}}]{Wilson:2000}
{Wilson}, T.~L., {Gaume}, R.~A., {Gensheimer}, P., \& {Johnston}, K.~J. 2000,
  \apj, 538, 665

\bibitem[{{Wright} {et~al.}(1996){Wright}, {Plambeck}, \&
  {Wilner}}]{Wright:1996}
{Wright}, M.~C.~H., {Plambeck}, R.~L., \& {Wilner}, D.~J. 1996, \apj, 469, 216

\bibitem[{{Zapata} {et~al.}(2011){Zapata}, {Loinard}, {Schmid-Burgk},
  {Rodr{\'{\i}}guez}, {Ho}, \& {Patel}}]{Zapata:2010a}
{Zapata}, L.~A., {Loinard}, L., {Schmid-Burgk}, J., {et~al.} 2011, \apjl, 726,
  L12

\bibitem[{{Zapata} {et~al.}(2009){Zapata}, {Schmid-Burgk}, {Ho},
  {Rodr{\'{\i}}guez}, \& {Menten}}]{Zapata:2009}
{Zapata}, L.~A., {Schmid-Burgk}, J., {Ho}, P.~T.~P., {Rodr{\'{\i}}guez}, L.~F.,
  \& {Menten}, K.~M. 2009, \apjl, 704, L45

\end{thebibliography}


\Online

\begin{appendix} 
\section{ALMA spectrum towards the ethylene glycol peak.\label{appendix}}

 \begin{figure*}
   \centering
\includegraphics[angle=-90,width=18cm, clip]{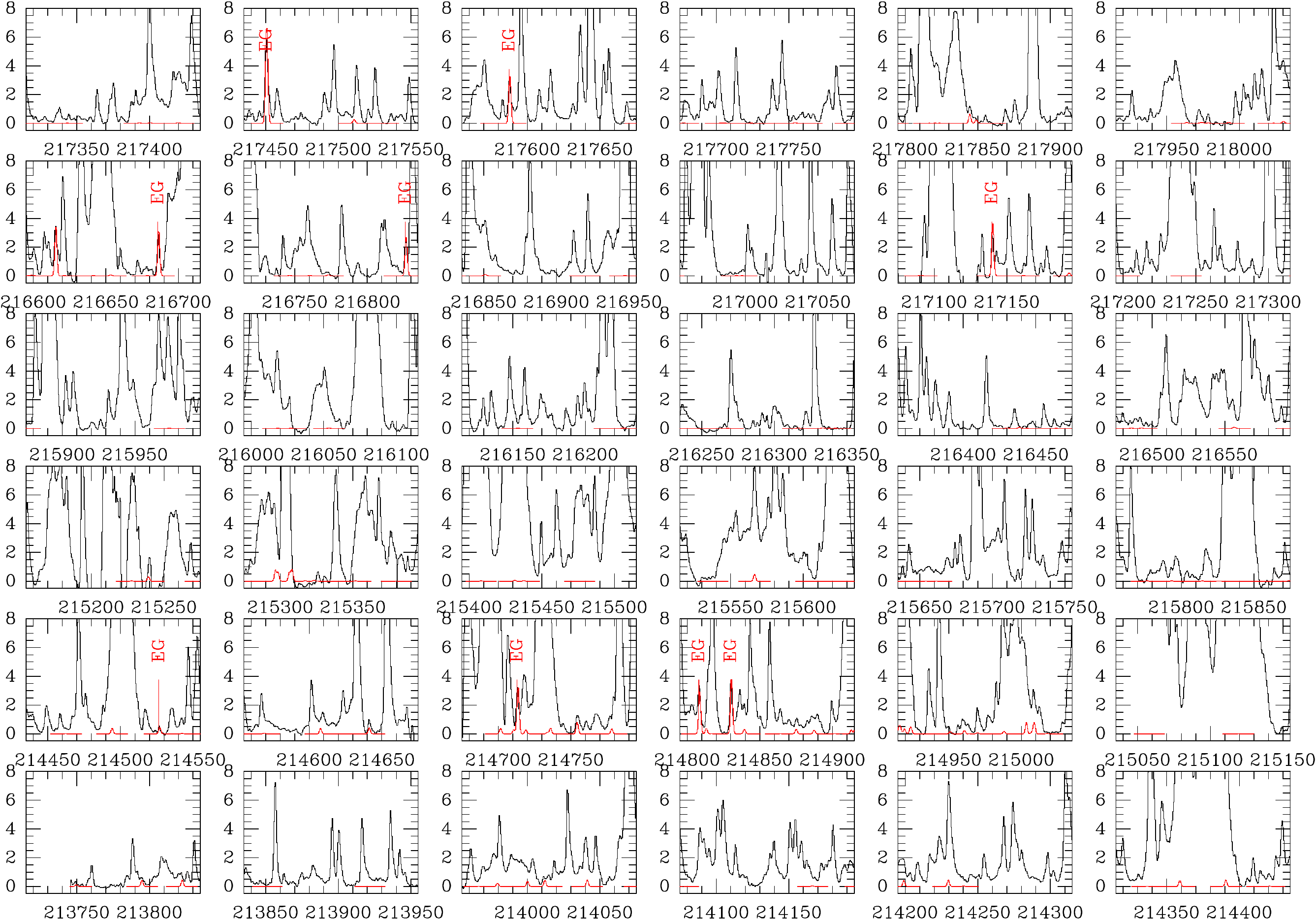}  
   \caption{ALMA spectrum (in black) towards the ethylene glycol peak. The synthetic spectrum is overlaid in red. The parameters used for the synthetic spectrum are: \textit{v}$\rm_{LSR}$ = 7.6~km~s$^{-1}$, $\Delta$\textit{v}$\rm_{1/2}$ = 2.3~km~s$^{-1}$, \textit{N}$\rm_{\textit{aGg'}}$ = 4.6~10$^{15}~$cm$^{-2}$, \textit{T} = 145~K. Each box corresponds to a 120~MHz bandwidth and the ordinates are the intensities in K. The frequencies (in MHz) increase from left to right and from bottom to top. The transitions listed in Table  \ref{table.freq-eg} are indicated on the plots by "EG".}
         \label{spesyn-all}
\end{figure*}
 \begin{figure*}
 \includegraphics[angle=-90,width=18cm, clip]{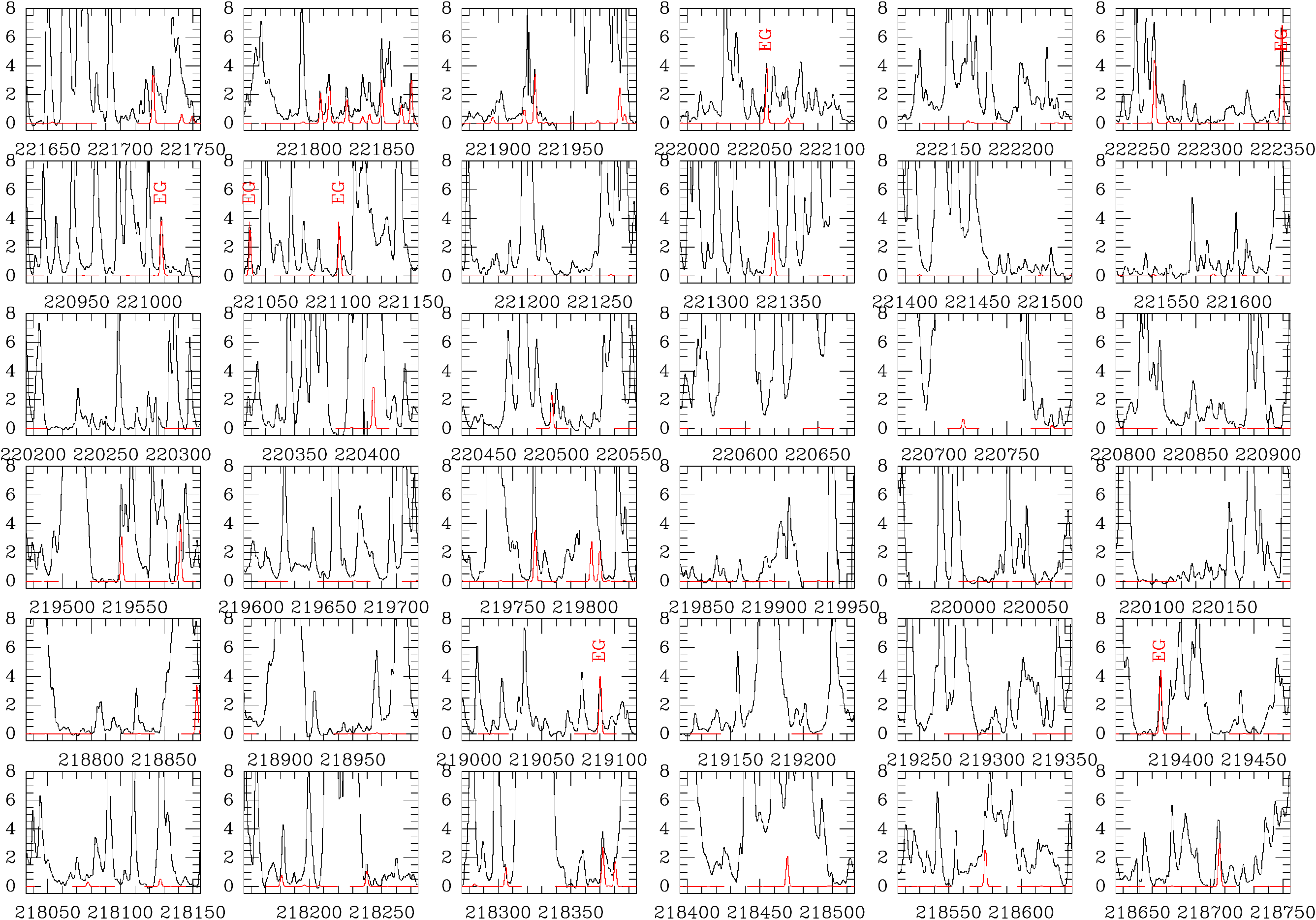} 
  \end{figure*}
   \begin{figure*}
\includegraphics[angle=-90,width=18cm, clip]{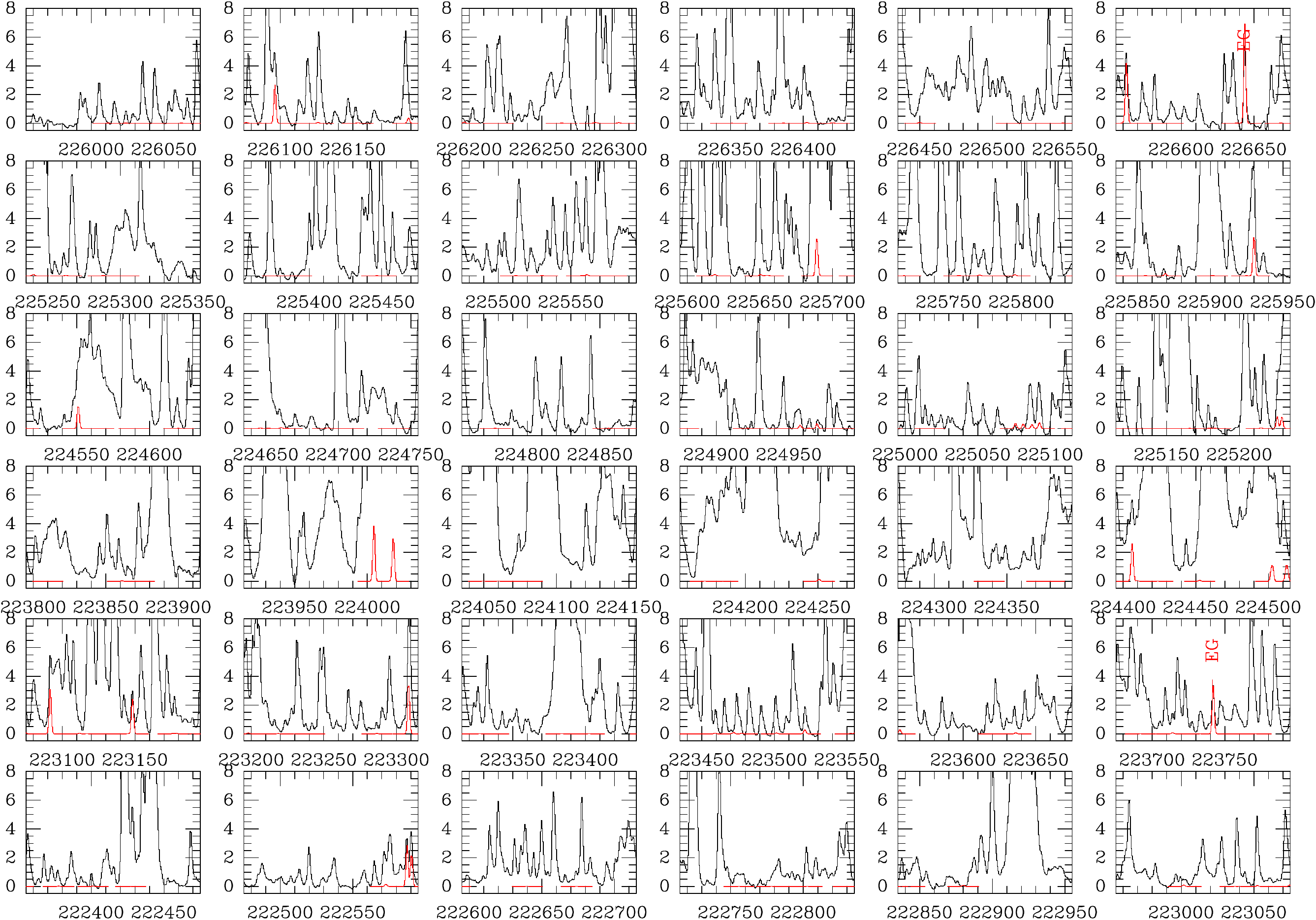} 
 \end{figure*}
  \begin{figure*}
\includegraphics[angle=-90,width=18cm, clip]{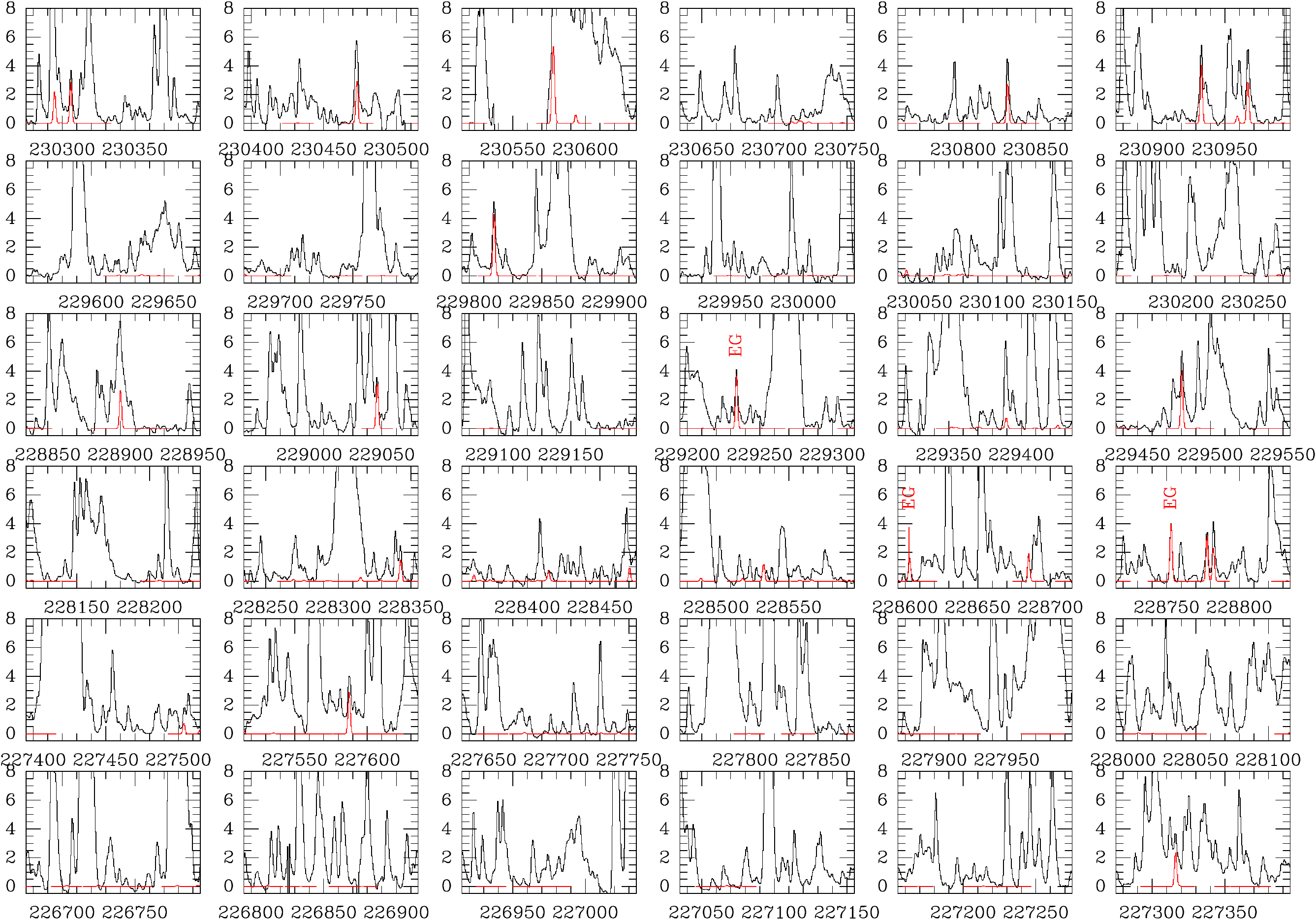} 
 \end{figure*}
  \begin{figure*}
\includegraphics[angle=-90,width=18cm, clip]{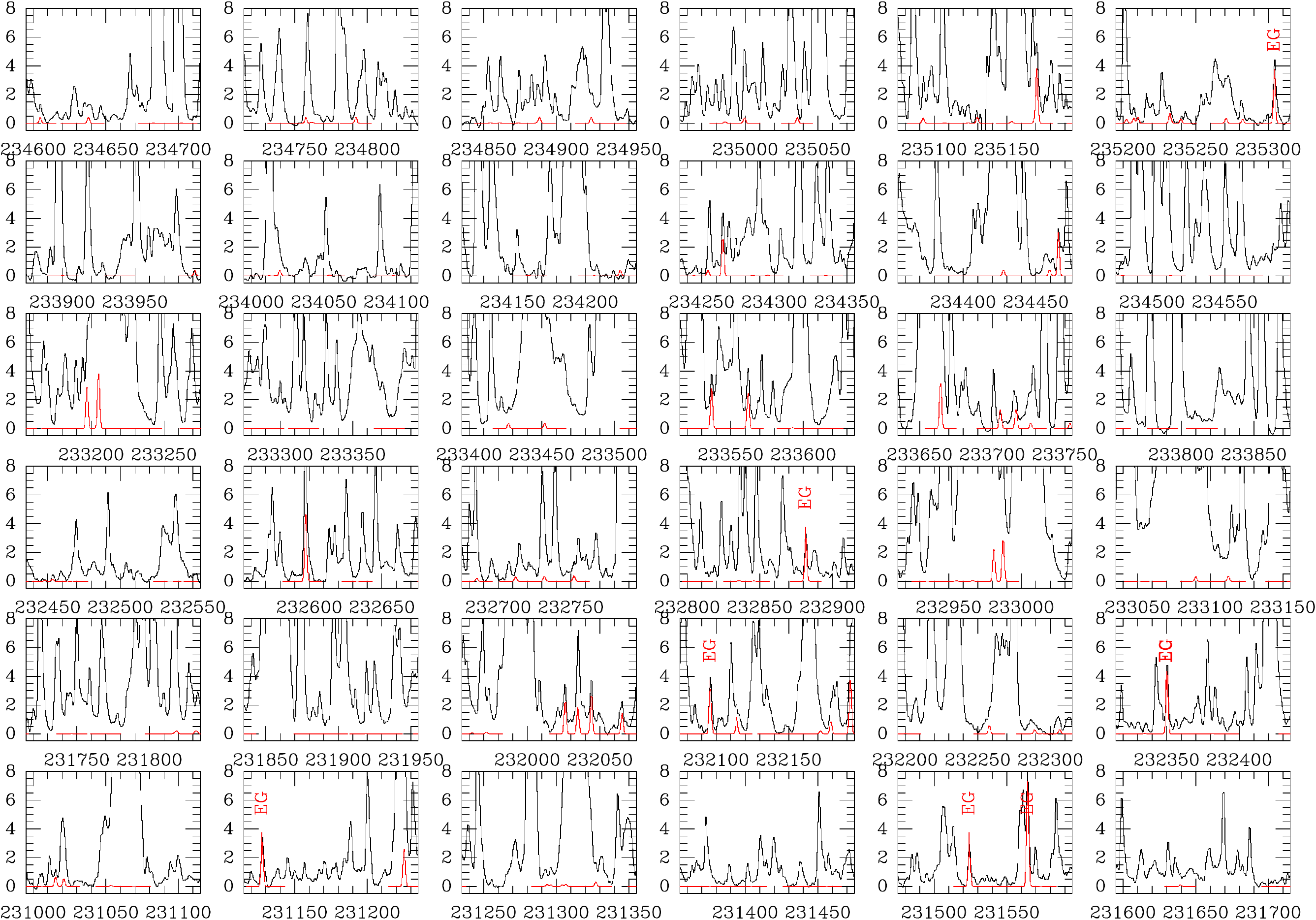} 
 \end{figure*}
  \begin{figure*}
\includegraphics[angle=-90,width=18cm, clip]{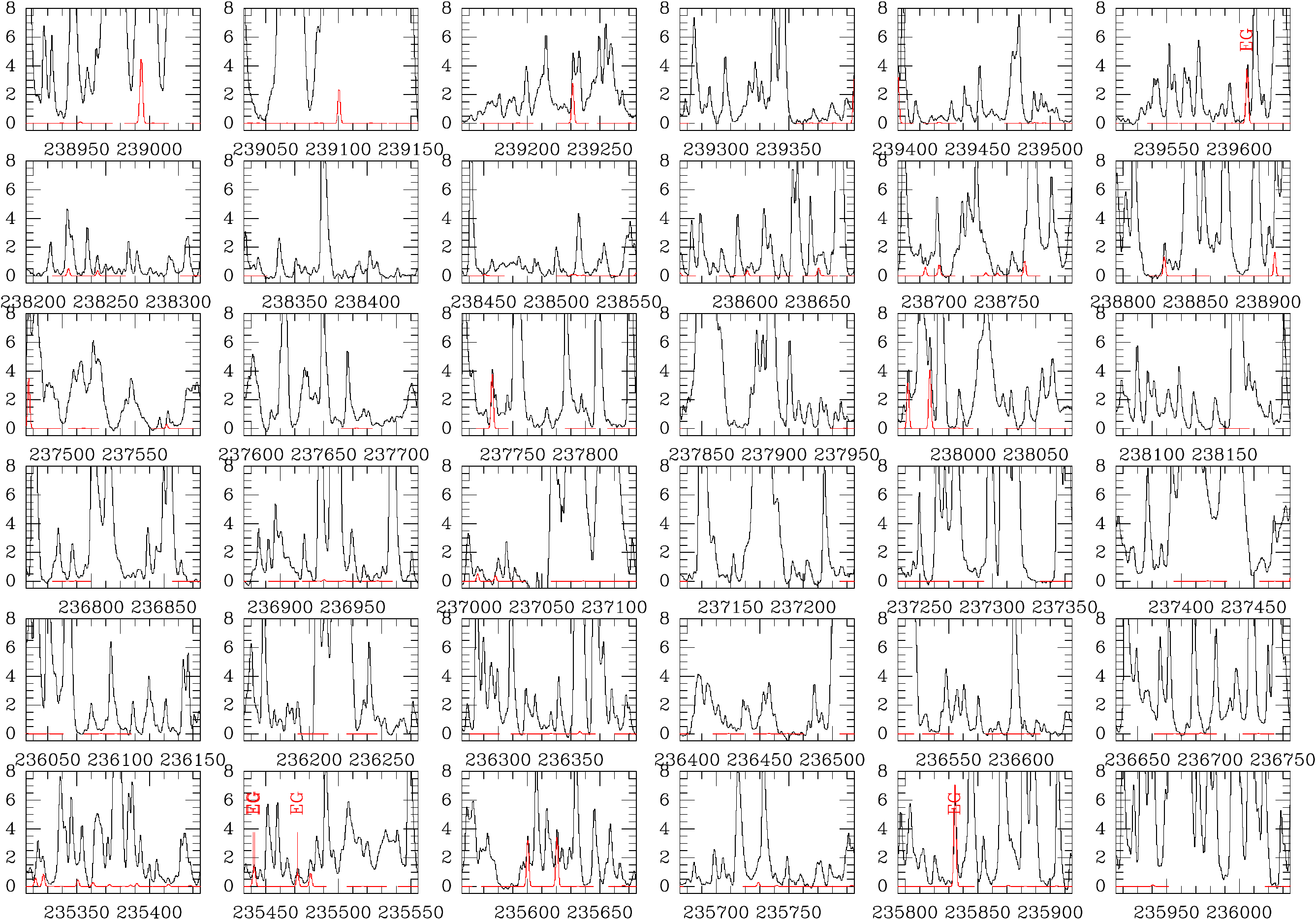} 
 \end{figure*}
  \begin{figure*}
\includegraphics[angle=-90,width=18cm, clip]{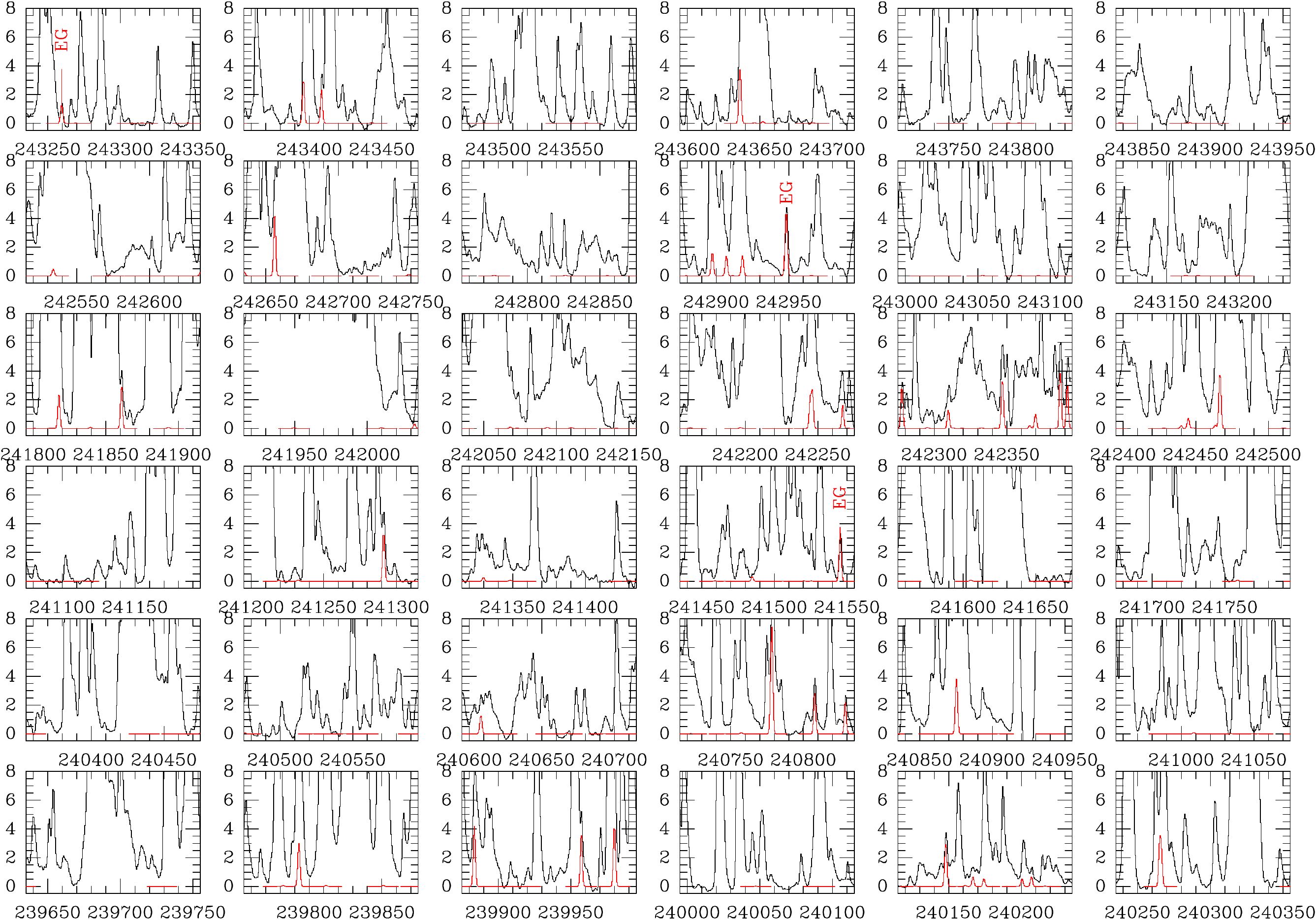} 
 \end{figure*}
  \begin{figure*}
\includegraphics[angle=-90]{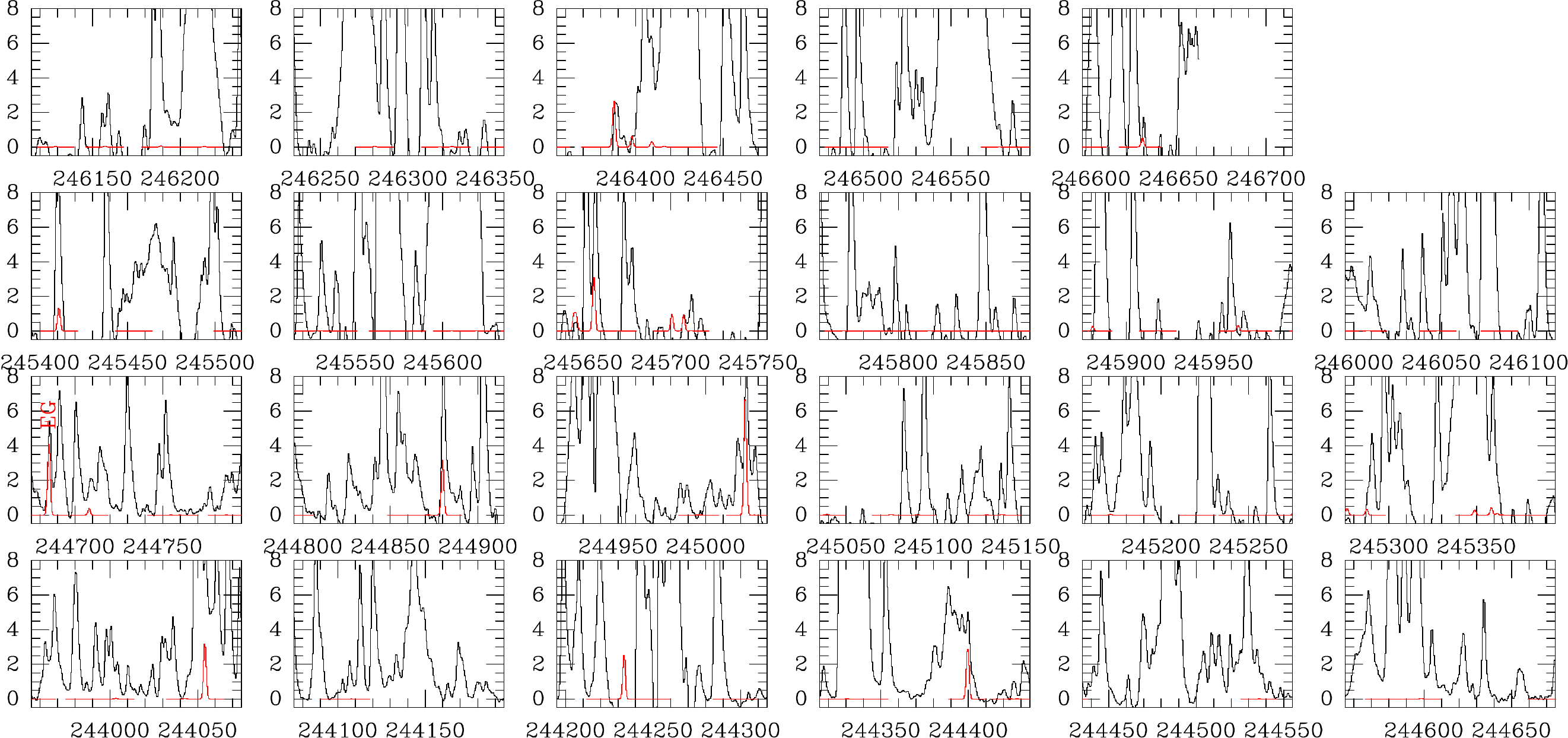} 
 \end{figure*}
 
\end{appendix}

\end{document}